\documentclass[seceq,preprint]{ptptex}

\usepackage{graphicx}

\usepackage{color}

\newcommand{\beq}{\begin{equation}}
\newcommand{\eeq}{\end{equation}}
\newcommand{\beqn}{\begin{eqnarray}}
\newcommand{\eeqn}{\end{eqnarray}}

\preprintnumber[3cm]{
OU-HET-670-2010\\TU-869}

\title{
Flavor Structure of the Three Site Higgsless Model 
}

\author{
Masafumi \textsc{Kurachi}$^1$
\ and \ 
Tetsuya \textsc{Onogi}$^2$
}

\inst{
$^1$Department of Physics, Tohoku University, Sendai 980-8578, Japan, \\
$^2$Department of Physics, Osaka University, Toyonaka 560-0043, Japan
}

\abst
{

We study the flavor structure in the three 
site Higgsless model. 
In this model, the gauge bosons and fermions have 
heavy partners, coming from the Kaluza-Klein excitation in the
dimensional deconstruction picture. The yukawa couplings 
are introduced in a way to minimize the flavor chaning neutral current
in the light sector at the tree level.
Due to the flavor mixing between the light and the heavy partner fields, 
new effects on FCNC's appear at
one-loop level. As an example of such FCNC
processes,  
we calculate the contribution to the 
$b \rightarrow s \gamma$ amplitude in the three site Higgsless model. 
Interestingly, heavy particles which exist in the three site Higgsless 
model do not completely decouple in the heavy-mass limit. One-loop level 
$b \rightarrow s \gamma$ amplitude is calculated by considering  
all possible combinations of particles in the loop, then it is compared 
to the experiment.
The result shows that the central value of the $B \rightarrow X_s \gamma$ 
branching ratio in the three site Higgsless model takes closer value 
to its experimental central value as one takes the larger value of 
a free parameter, $\varepsilon_{tR}$, within a range allowed 
by the precision electroweak measurement. 
}

\begin{document}

\maketitle
\tableofcontents

\vspace{1cm}
\section{Introduction}

The three site Higgsless model~\cite{SekharChivukula:2006cg} has 
been proposed as a low energy effective theoy of a
class of models which realize an alternative electroweak symmetry
breaking scenario to the Standard Model (SM). 
Unlike the SM, the three site Higgsless model 
does not have any elementary scalar particle as a low energy degree 
of freedom, instead, it possesses vector resonances, which can be 
viewed as heavy partners of the W and Z bosons. Those heavy 
gauge bosons contribute to suppress the
longtudinal weak gauge boson scattering amplitudes and make them
satisfy the perturbative unitarity bound at higher energy region 
in a similar fashion as in the case of usual Higgsless models.
\cite{Csaki:2003dt,Csaki:2003zu,SekharChivukula:2001hz, 
Chivukula:2002ej,Chivukula:2003kq}
In fact, the Higgsless models has been
studied~\cite{decHless1,precision, ideal,decHless2}
using the  
technique of dimensional deconstruction~\cite{ArkaniHamed:2001ca, 
Hill:2000mu}. However, if the heavy gauge bosons
are too light, they can give measurable effects. 
For this reason, constraints from precision electroweak 
measurements~\cite{Peskin-Takeuchi} have been extensively 
investigated~\cite{precision}, and it was shown that
the Higgsless model can satisfy
perturbative unitariy bounds at TeV scale but yet provide acceptably
small precision 
electroweak correction only when light fermions are ``ideally 
delocalized"~\cite{ideal}. It was also pointed out that consistency with 
multi-gauge-boson coupling measurement put lower bound on the 
mass of $W'$ and $Z'$ gauge bosons~\cite{Chivukula:2005ji}.

The three site Higgsless model can be viewed as an extremely deconstructed 
version of five-dimensional Higgsless model. It contains sufficient 
complexity to incorporate interesting physics, 
such as ideal fermion delocalization and the generation of fermion 
masses (including the top quark mass), yet remains simple enough 
for a practical study of hadron collider 
phenomenology. (See, for example, Ref.~\cite{He:2007ge}.)

In this paper, we study the flavor structure of the three site
Higgsless model.  
As will be shown in Section~\ref{overview}, light
quarks have almost flavor universal wavefunctions on the three sites.
Because of this feature, the light fermions behave in a very similar
way as those in the Standard Model (SM). However, this is not the case
for the top quark. Having a large mass, the top quark has distinct
wavefunction on the three sites and the interaction obtained from the
wavefunction overlap can have nonuniversal flavor structure.
Therefore, flavor physics which involve third 
generation fermions have a potential to 
distinguish the three site Higgsless model and the SM. As an example of 
such processes, we investigate $b \rightarrow s \gamma$ decay rate 
in the three site Higgsless model, and compare it to the SM prediction and 
experimental results. It will be shown that the central value of the $B \rightarrow X_s \gamma$ 
branching ratio in the three site Higgsless model takes closer value 
to its experimental central value as one takes the larger value of 
the model parameter, $\varepsilon_{tR}$, within a range allowed 
by the precision electroweak measurement.

The paper is organized as follows. After reviewing the 
three site Higgsless model in the next section, we introduce the generation 
structure to the three site Higgsless model in Section \ref{sec:flavor_structure}. 
Forms of fermion couplings to gauge/Nambu-Goldstone (NG) bosons are also 
given in Section \ref{sec:flavor_structure}. 
Section \ref{sec:bsgamma} is devoted to the study of $b \rightarrow s \gamma$ 
process in the three site Higgsless model. Concluding remarks can be 
found in Section \ref{sec:conclusions}. Detailed explanations of the three site 
Higgsless model, and expressions of couplings are summarized in Appendices.

\vspace{1cm}
\section{Three Site Higgsless Model: An Overview}\label{overview}

In this section, we give an overview of the three site Higgsless model which 
was proposed in Ref.~\cite{SekharChivukula:2006cg}.  
First, we review the electroweak gauge sector of the 
three site Higgsless model, then the fermion sector of the model in the 
case of one fermion generation is explained.

\begin{figure}[t]
\begin{center}
  \includegraphics[width=0.5\textwidth]{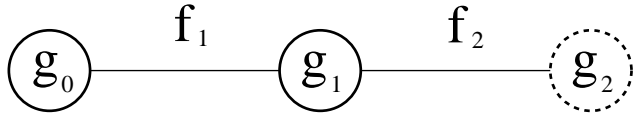}
\end{center}
\caption{Graphical expression of the electroweak gauge sector of 
the three site Higgsless model. Circles placed at ``sites" represent 
gauge symmetries, while ``links" between sites represent the non-linear 
sigma fields. In this paper, we call these sites, from left to right, as site-0, site-1, and site-2, 
respectively. }
\label{fig:threesite}
\end{figure}

The gauge sector of the three site Higgsless model can be expressed 
by using so-called  ``Moose notation'' \cite{Georgi:1985hf} as 
illustrated in Fig.~\ref{fig:threesite}.  
The model incorporates an
$SU(2) \times SU(2) \times U(1)$ gauge group 
(represented by circles with gauge couplings $g_0$, $g_1$, and $g_2$, respectively), 
and two nonlinear 
$(SU(2)\times SU(2))/SU(2)$ sigma fields 
(represented by links with decay constants $f_1$ and $f_2$) 
in which the global symmetry 
groups in adjacent sigma fields are identified with the corresponding 
factors of the gauge group.  The symmetry breaking between the middle 
$SU(2)$ and the $U(1)$ follows an $SU(2)_L \times SU(2)_R/SU(2)_V$ symmetry 
breaking pattern with the $U(1)$ embedded as the $T_3$-generator of $SU(2)_R$. 
Thus, the Lagrangian of the gauge sector of the three site Higgsless model 
is given by
\beq
{\cal L} = \sum_{j=0}^{2} -\frac{1}{2} {\rm Tr} \left( F_{j \mu\nu}F_j^{\mu\nu} \right)
+ \sum_{j=1}^{2} \frac{f_j^2}{4} {\rm Tr} \left[ \left(D_\mu U_j\right)^\dagger \left(D^\mu U_j\right)\right],
\label{eq:L}
\eeq
with
\beq
D_\mu U_j = \partial_\mu -i A_\mu^{j-1} U_j + i U_j A_\mu^{j},
\eeq
where the first two gauge fields $(j=0, 1)$ correspond to those of 
$SU(2)$ gauge groups, while the last 
gauge fields $(j=2)$ corresponds to that of the $U(1)$ gauge group. 
Non-linear sigma model fields are defined by NG bosons $\pi_j \equiv \pi_j^a T^a$ 
(where $T^a$ is the generator of the $SU(2)$ gauge group) 
and decay constants $f_j$ as
\beq
U_j \equiv \exp{\left[ 2 i \pi_j/f_j \right]}.
\eeq
In the three site Higgsless model, it is assumed that both of two nonlinear sigma 
fields have the same decay constant, $f=\sqrt{2}v$ ($v\simeq 246$GeV), 
for simplicity (and also to achieve maximal delay of the energy scale 
of perturbative unitarity violation). 
Also, $g_1 \gg g_0, g_2$ is assumed to obtain the 
hierarchy between Standard Model (SM) 
gauge bosons and their heavy partners $W'$ and $Z'$. 

For the purpose of calculating $b \rightarrow s \gamma$ amplitude, it
is convenient 
to work in the mass eigenbasis. In Appendix \ref{app:gauge}, we summarize the mass 
spectrum and the forms of mass eigenfunctions of the gauge
bosons. A detailed 
discussion on Nambu-Goldstone (NG) boson is also given there. 
Appendix \ref{app:GNGcouplings} summarizes expressions of couplings among gauge/NG bosons 
which will be needed for the calculation of the $b \rightarrow s \gamma$ amplitude.

The gauge sector of the three site Higgsless model explained above is
the same as that  
in models of extended  electroweak gauge symmetries 
\cite{Casalbuoni:1985kq,Casalbuoni:1995qt} as the
hidden local symmetries~\cite{Bando:1985ej,Bando:1985rf,
Bando:1988ym,Bando:1988br,Harada:2003jx}
in the 
electroweak symmetry breaking sector.  The new
physics discussed in Ref.~\cite{SekharChivukula:2006cg}  
relates to the fermon sector, which we briefly review below.

In the three site Higgsless model, two left-handed fermions and 
two right-handed fermions are introduced for each fermion flavor. 
In the case of up (down) quark for example, two left-handed 
fermions are denoted as $u_{L0}, u_{L1}$ ($d_{L0}, d_{L1} $), while 
two right-handed 
fermions are denoted as $u_{R1}, u_{R2}$  $(d_{R1}, d_{R2}) $. 
$u_{L0}$ and $d_{L0}$ form a doublet of SU(2) at site-0, and denoted as 
$\psi_{L0}\equiv(u_{L0}, d_{L0})^T$. In the same way, $u_{L1}$ and $d_{L1}$, as 
well as $u_{R1}$ and $d_{R1}$ form doublets of SU(2) at site-1, and denoted as 
$\psi_{L1}\equiv(u_{L1}, d_{L1})^T$ and $\psi_{R1}\equiv(u_{R1}, d_{R1})^T$, respectively. 
$u_{2R}$ and $d_{2R}$ are SU(2) singlet, and have U(1) charge at site-2 
($+2/3$ and $-1/3$, respectively). In the three site Higgsless model, it is assumed that 
doublets also have U(1) charge at site-2, and those are taken to be 
$+1/6$ for all doublets.\footnote{This U(1) charge assignment causes   
non-local interactions when we consider the ``continuum limit" of the model, 
however, it can be easily extended to ``four-site" by adding an 
extra U(1) site to cure the problem. (See Ref.~\cite{Belyaev:2009ve}.)
What we should emphasize here is  that this U(1) charge assignment is 
perfectly consistent as a four-dimensional gauged non-linear sigma model.
}

With these assignments, we may write the Yukawa couplings and quark mass term
as
\beq
{\cal L}_{q} = 
- \sqrt{2}v \lambda \bar{\psi}_{L0} U_1 \psi_{R1}
   -  \sqrt{2}v  \tilde{\lambda} \bar{\psi}_{R1} \psi_{L1} 
   -  \sqrt{2}v \bar{\psi}_{L1}  U_2 
         \left( 
           \begin{array}{cc}
              \lambda_u' & \\
              & \lambda_d'
           \end{array}
         \right)
         \left( 
           \begin{array}{c}
              u_{R2} \\
              d_{R2}
           \end{array}
         \right) + h.c. 
\eeq
Then the quark mass terms are expressed as
\beqn
  {\cal L}_{mass} &=&
    - \sqrt{2}v \left( 
           \begin{array}{cc}
              \bar{u}_{L0} & 
              \bar{d}_{L0} 
           \end{array}
         \right)
         \left( 
           \begin{array}{cc}
              \lambda & \\
              & \lambda
           \end{array}
         \right)
         \left( 
           \begin{array}{c}
              u_{R1} \\
              d_{R1}
           \end{array}
         \right)
\nonumber\\
  &&
    - \sqrt{2}v
         \left( 
           \begin{array}{cc}
              \bar{u}_{R1} & 
              \bar{d}_{R1} 
           \end{array}
         \right)
         \left( 
           \begin{array}{cc}
              \tilde{\lambda} & \\
              & \tilde{\lambda}
           \end{array}
         \right)
         \left( 
           \begin{array}{c}
              u_{L1} \\
              d_{L1}
           \end{array}
         \right)
\nonumber\\
  &&
    - \sqrt{2}v
         \left( 
           \begin{array}{cc}
              \bar{u}_{L1} & 
              \bar{d}_{L1} 
           \end{array}
         \right)
         \left( 
           \begin{array}{cc}
              \lambda_u' & \\
              & \lambda_d'
           \end{array}
         \right)
         \left( 
           \begin{array}{c}
              u_{R2} \\
              d_{R2}
           \end{array}
         \right) + h.c.
\eeqn
Thus, the fermion mass matrix for a specific flavor, say $f$, takes the following form:
\begin{equation}
M_{f} = \sqrt{2} v
 \left(
 \begin{array}{cc}
 \lambda & 0\\
 \tilde{\lambda} & \lambda'_f
 \end{array}
\right)
\equiv
 M
 \left(
 \begin{array}{cc}
 \varepsilon_L & 0\\
 1 & \varepsilon_{fR}
 \end{array}
\right).
\end{equation}
In the last expression, we factored out the overall scale $\sqrt{2}v\tilde{\lambda} (\equiv M)$, 
by introducing flavor-universal parameter $\varepsilon_L(\equiv \lambda/\tilde{\lambda})$ 
and flavor-dependent parameter $\varepsilon_{fR}(\equiv \lambda'_f/\tilde{\lambda})$.
After diagonalizing the above expression, one obtains two Dirac fermions for 
each flavor: the lighter one, $f (= u, d, c, s, t, b)$, is identified as SM fermion, 
while the other, $F (= U, D, C, S, T, B)$, is a heavy 
partner of it, which does not exist in the SM. Expressions of masses and 
wavefunctions of them are obtained by diagonalizing the above mass matrix 
perturbatively in $\varepsilon_L$:~\cite{SekharChivukula:2006cg}
\begin{equation}
m_f  = {M \varepsilon_L \varepsilon_{fR} \over \sqrt{1+ \varepsilon^2_{fR}}}
\left[
1 - {\varepsilon^2_L \over 2\,(\varepsilon^2_{fR}+1)^2}+\ldots
\right]~,\\
\label{eq:fermion_mass}
\end{equation}
\begin{align}
f_L & = f^0_L\, \psi^f_{L0} + f^1_L\,\psi^f_{L1} \nonumber \\
& = \left(-1 + {\varepsilon^2_L \over 2(1+\varepsilon^2_{fR})^2} +\ldots \right) \psi^f_{L0}
+\left({\varepsilon_L \over 1+\varepsilon^2_{fR}} + 
{(2\varepsilon^2_{fR}-1)\varepsilon^3_L\over 2(\varepsilon^2_{fR} + 1)^3} + \ldots\right) \psi^f_{L1} 
\\
f_R & = f^1_R\, \psi^f_{R1} + f^2_R\,f_{R2} \nonumber \\
& = \left(-\,{\varepsilon_{fR} \over \sqrt{1+\varepsilon^2_{fR}}} + 
{\varepsilon_{fR}\, \varepsilon^2_L \over (1+\varepsilon^2_{fR})^{5/2}}+\ldots\right)\psi^f_{R1}
+\left({1\over \sqrt{1+\varepsilon^2_{fR}}}+{\varepsilon^2_{fR}\,\varepsilon^2_L \over
(1+\varepsilon^2_{fR})^{5/2}}+\ldots\right)f_{R2}~,
\end{align}
\begin{equation}
m_F = M \sqrt{1+\varepsilon^2_{fR}}
\left[
1+ {\varepsilon^2_L \over 2\,(\varepsilon^2_{fR}+1)^2}+\ldots
\right]~,
\end{equation}
\begin{align}
F_L & = F^0_L \psi^f_{L0} + F^1_L \psi^f_{L1} \nonumber\\
& = \left(-\,{\varepsilon_L \over 1+\varepsilon^2_{fR}} - 
{(2\varepsilon^2_{fR}-1)\varepsilon^3_L\over 2(\varepsilon^2_{fR} + 1)^3} + \ldots\right) \psi^f_{L0} 
+ \left(-1 + {\varepsilon^2_L \over 2(1+\varepsilon^2_{fR})^2} +\ldots \right) \psi^t_{L1}\\
F_R & = F^1_R \psi^f_{R1} + F^2_R f_{R2}~,\nonumber\\
&= \left(-\,{1\over \sqrt{1+\varepsilon^2_{fR}}}-{\varepsilon^2_{fR}\,\varepsilon^2_L \over
(1+\varepsilon^2_{fR})^{5/2}}+\ldots\right)\psi^f_{R1} +
\left(-\,{\varepsilon_{fR} \over \sqrt{1+\varepsilon^2_{fR}}} + 
{\varepsilon_{fR}\, \varepsilon^2_L \over (1+\varepsilon^2_{fR})^{5/2}}+\ldots\right)f_{R2}~,
\end{align}
Though $\varepsilon_L$ is a priori a free parameter which is not necessarily small, 
in the three site Higgsless model, it is taken to be proportional to 
$x  \simeq 2\frac{M_W}{M_{W'}} (\ll 1)$:
\begin{equation}
\varepsilon_L^2 = \frac{x^2}{2}+\frac{x^4}{8} + \cdots.
\label{eq:ideal}
\end{equation}
This is a choice for ``ideal fermion delocalization" with which precision electroweak 
corrections are minimized~\cite{ideal}.\footnote{
Since the expression in Eq.~(\ref{eq:ideal}) was obtained by tree-level analysis, 
it needs to be modified if we include loop corrections. Actually, in Ref.~\cite{Abe:2008hb}, 
it was pointed out that the tree-level ideal delocalization as in Eq.~(\ref{eq:ideal})  
is ruled out at 95\% CL if large bosonic loop corrections are included. 
(See Fig. 2-4 in Ref.~\cite{Abe:2008hb}.) Such one-loop correction to the delocalization 
parameter $\varepsilon_L$ could affect the leading order calculation 
of the $b \rightarrow s \gamma$ amplitude, however, the effect is numerically small 
and within the uncertainty of the current calculation. Thus, in this paper,  
we use the tree-level ideal delocalization for calculational simplicity.}
Thus the perturbative expansion in $\varepsilon_L$ in 
expressing masses and wavefunctions is justified in the three site Higgsless model.

As for $\varepsilon_{tR}$, it was
shown~\cite{SekharChivukula:2006cg,Abe:2008hb} that a value large than
$\varepsilon_{tR} \sim 0.3$ would cause large isospin violation, which
is inconsistent with precision electroweak measurement. Thus, in
the numerical study of the present paper, we only consider the range
of $0 \le 
\varepsilon_{tR} \le 0.3$. 
It is also worth mentioning that $\varepsilon_{fR}$  
for quarks other than the top is negligibly small since $\varepsilon_{fR}$ 
for each flavor is almost proportional to its mass. Therefore, in the numerical 
study, we take $\varepsilon_{fR} = 0$ for light quarks unless it appears as 
leading contribution.

\vspace{15mm}
\section{Flavor Structure of the Three Site Higgsless Model}
\label{sec:flavor_structure}
In the previous section, we explained the Lagrangian of the fermion sector 
with ignoring the mixings among different flavors. Here, we introduce the 
generation mixing matrix to the three site Higgsless model in a minimal way, 
and show how fermion mass eigenstates couple to gauge/NG bosons.

\subsection{Introducing the generation mixing matrix}
It is straightforward to incorporate flavor  
mixing in a minimal way. Adding generational indices 
to each of the fermion fileds, we may choose  
$\lambda$ and $\tilde{\lambda}$ to be proportional 
to unit matrix in order to keep the flavor mixing as
minimal as possible. $\lambda_u'$ and $\lambda_d'$ are 
generalized to $3\times 3$ general complex-valued matrices 
$(\Lambda_u')_{IJ}$ and $(\Lambda_d')_{IJ}$ where 
$I, J\ (= 1 \sim 3)$ are the generational indices.
\footnote{
This way of introducing flavor structure was briefly mentioned 
in Ref.\cite{SekharChivukula:2006cg}. Here, we give detailed explanation 
and derive explicit forms of fermion couplings to gauge/NG bosons 
with including flavor mixings.
}
\begin{eqnarray}
{\cal L}_{mass} 
  &=&
     -\sqrt{2}v \left( 
           \begin{array}{cc}
              \bar{u}_{L(0,I)} & 
              \bar{d}_{L(0,I)}
           \end{array}
         \right)
         \left( 
           \begin{array}{c|c}
              \lambda \cdot \delta_{IJ} & \\ \hline
              & \lambda \cdot \delta_{IJ}
           \end{array}
         \right)
         \left( 
           \begin{array}{c}
              u_{R(1,J)} \\
              d_{R(1,J)}
           \end{array}
         \right)
\nonumber\\
  &&
    - \sqrt{2}v
         \left( 
           \begin{array}{cc}
              \bar{u}_{R(1,I)} & 
              \bar{d}_{R(1,I)} 
           \end{array}
         \right)
         \left( 
           \begin{array}{c|c}
              \tilde{\lambda} \cdot \delta_{IJ} & \\ \hline
              & \tilde{\lambda} \cdot \delta_{IJ}
           \end{array}
         \right)
         \left( 
           \begin{array}{c}
              u_{L(1,J)} \\
              d_{L(1,J)}
           \end{array}
         \right)
\nonumber\\
  &&
    - \sqrt{2}v
         \left( 
           \begin{array}{cc}
              \bar{u}_{L(1,I)} & 
              \bar{d}_{L(1,I)} 
           \end{array}
         \right)
         \left( 
           \begin{array}{c|c}
              (\Lambda_u')_{IJ} & \\ \hline
              & (\Lambda_d')_{IJ}
           \end{array}
         \right)
         \left( 
           \begin{array}{c}
              u_{R(2,J)} \\
              d_{R(2,J)}
           \end{array}
         \right) + h.c.
\end{eqnarray}
Here, $u_{L(i,I)}$ etc. represent fermion fields at $i$th site with generation index $I$.
Using the Hermitian matrices $V_{uL}, V_{uR}, V_{dL}, V_{dR}$ which 
diagonalize $(\Lambda_u')_{IJ}$ and $(\Lambda_d')_{IJ}$, 
\begin{eqnarray}
V_{uL} \Lambda_u' V_{uR}^\dagger 
&=& 
         \left( 
           \begin{array}{ccc}
             \lambda_u^{'(1)}&& \\
             &\lambda_u^{'(2)}& \\
             &&\lambda_u^{'(3)} \\
           \end{array}
         \right), \\
V_{dL} \Lambda_d' V_{dR}^\dagger 
&=& 
         \left( 
           \begin{array}{ccc}
             \lambda_d^{'(1)}&& \\
             &\lambda_d^{'(2)}& \\
             &&\lambda_d^{'(3)} \\
           \end{array}
         \right),
\end{eqnarray}
we define the new base of quark fields as
\begin{eqnarray}
u_{L(0,I')}' &=& \sum_{I=1}^{3}\ (V_{uL})^{I'I}\, u_{L(0,I)},  
\ \ \ \ \ d_{L(0,I')}' = \sum_{I=1}^{3}\ (V_{dL})^{I'I}\, d_{L(0,I)},\label{eq:base1}\\
u_{R(1,I')}' &=& \sum_{I=1}^{3}\ (V_{uL})^{I'I}\, u_{R(1,I)},
\ \ \ \ \ d_{R(1,I')}' = \sum_{I=1}^{3}\ (V_{dL})^{I'I}\, d_{R(1,I)},\\
u_{L(1,I')}' &=& \sum_{I=1}^{3}\ (V_{uL})^{I'I}\, u_{L(1,I)},
\ \ \ \ \ d_{L(1,I')}' = \sum_{I=1}^{3}\ (V_{dL})^{I'I}\, d_{L(1,I)},\\
u_{R(2,I')}' &=& \sum_{I=1}^{3}\ (V_{uR})^{I'I}\, u_{R(2,I)},
\ \ \ \ \ d_{R(2,I')}' = \sum_{I=1}^{3}\ (V_{dR})^{I'I}\, d_{R(2,I)}.\label{eq:base2}
\end{eqnarray}
With this base, the fermion mass terms become diagonal in generation-index space:
\begin{eqnarray}
{\cal L}_{mass} 
  &=&
    - \sqrt{2}v \left( 
           \begin{array}{cc}
              \bar{u}_{L(0,I')}' & 
              \bar{d}_{L(0,I')}'
           \end{array}
         \right)
         \left( 
           \begin{array}{c|c}
              \lambda \cdot \delta_{I'J'} & \\ \hline
              & \lambda \cdot \delta_{I'J'}
           \end{array}
         \right)
         \left( 
           \begin{array}{c}
              u_{R(1,J')}' \\
              d_{R(1,J')}'
           \end{array}
         \right)
\nonumber\\
  &&
    - \sqrt{2}v
         \left( 
           \begin{array}{cc}
              \bar{u}_{R(1,I')}' & 
              \bar{d}_{R(1,I')}' 
           \end{array}
         \right)
         \left( 
           \begin{array}{c|c}
              \tilde{\lambda} \cdot \delta_{I'J'} & \\ \hline
              & \tilde{\lambda} \cdot \delta_{I'J'}
           \end{array}
         \right)
         \left( 
           \begin{array}{c}
              u_{L(1,J')}' \\
              d_{L(1,J')}'
           \end{array}
         \right)
\nonumber\\
  &&
    - \sqrt{2}v
         \left( 
           \begin{array}{cc}
              \bar{u}_{L(1,I')}' & 
              \bar{d}_{L(1,I')}' 
           \end{array}
         \right)
         \left( 
           \begin{array}{c|c}
              \lambda_u'^{(I')}\cdot \delta_{I'J'} & \\ \hline
              & \lambda_d'^{(I')}\cdot \delta_{I'J'}
           \end{array}
         \right)
         \left( 
           \begin{array}{c}
              u_{R(2,J')}' \\
              d_{R(2,J')}'
           \end{array}
         \right) + h.c. \nonumber \\
\end{eqnarray}

For each generation index $I'$, we can further diagonalize the above mass matrix 
exactly in the same way as it was done for one-generation case in the previous 
section:\footnote{Note that the index $I'$ is not summed over here.}
\begin{eqnarray}
u_{L(i',I')}'' &=& \sum_{i=0}^{1}\ \left(U_{uL}^{(I')} \right)^{i'i}\, u_{L(i,I')}',  
\ \ \ \ \  d_{L(i',I')}'' = \sum_{i=0}^{1}\ \left(U_{dL}^{(I')} \right)^{i'i}\, d_{L(i,I')}',  
\nonumber \\
u_{R(i',I')}'' &=& \sum_{i=1}^{2}\ \left(U_{uR}^{(I')} \right)^{i'i}\, u_{R(i,I')}',  
\ \ \ \ \  d_{R(i',I')}'' = \sum_{i=1}^{2}\ \left(U_{dR}^{(I')} \right)^{i'i}\, d_{R(i,I')}'.
\label{eq:ftrans}
\end{eqnarray}
Here, $i' (=0, 1)$ represents the KK-mode index, and $U$'s are the rotation matrix 
made of eigenvectors of light and heavy mass eigenstates. 
Now $u_{L,R(i',I')}''$ and $d_{L,R(i',I')}''$ are mass eigenstates, thus we give specific names 
to them:
\begin{eqnarray}
u_{L,R(0,1)}'' &=& u_{L,R} \, ,  \ \ \ \ \,   d_{L,R(0,1)}'' = d_{L,R} \, ,  \\
u_{L,R(0,2)}'' &=& c_{L,R} \, ,  \ \ \ \ \   d_{L,R(0,2)}'' = s_{L,R} \, ,  \\
u_{L,R(0,3)}'' &=& t_{L,R} \, ,  \ \ \ \ \ \,  d_{L,R(0,3)}'' = b_{L,R} \, ,  \\
u_{L,R(1,1)}'' &=& U_{L,R} \, ,  \ \ \ \ \,  d_{L,R(1,1)}'' = D_{L,R} \, ,  \\
u_{L,R(1,2)}'' &=& C_{L,R} \, ,  \ \ \ \ \,  d_{L,R(1,2)}'' = S_{L,R} \, ,  \\
u_{L,R(1,3)}'' &=& T_{L,R} \, ,  \ \ \ \ \   d_{L,R(1,3)}'' = B_{L,R} \, ,
\end{eqnarray}
where $u, d, c, s, t, b$ are identified as the SM quarks, while 
$U, D, C, S, T, B$ are their heavy partners. One can
also introduce the flavor structure of the lepton sector in
a similar manner.

\vspace{5mm}
\subsection{Fermion couplings to neutral gauge bosons}
``$T_3$ part" of the fermion couplings to neutral gauge bosons in the
Lagrangian are 
expressed as 
\begin{eqnarray}
&& \sum_{i=0}^{1} \sum_{I=1}^{3}
\frac{g_i}{2} 
\left[
 \bar{u}_{L(i,I)} \gamma_\mu  u_{L(i,I)} - \bar{d}_{L(i,I)} \gamma_\mu  d_{L(i,I)}
\right] 
A_i^{3\, \mu}
\nonumber \\
&&\ \ \ \ \ \ \ \ \ \ \ +  
\sum_{i=1}^{2} \sum_{I=1}^{3} \frac{g_i}{2} 
\left[
 \bar{u}_{R(i,I)} \gamma_\mu  u_{R(i,I)} - \bar{d}_{R(i,I)} \gamma_\mu  d_{R(i,I)}
\right]
 A_i^{3\, \mu},
\end{eqnarray}
while hypercharge couplings are expressed as 
\begin{eqnarray}
&& \sum_{i=0}^{1} \sum_{I=1}^{3}
\frac{g_2}{6} 
\left[
 \bar{u}_{L(i,I)} \gamma_\mu  u_{L(i,I)} + \bar{d}_{L(i,I)} \gamma_\mu  d_{L(i,I)}
\right] 
A_2^{3\, \mu}
 \\
&& +  
\sum_{I=1}^{3} g_2 
\left[
\frac{1}{6}
\left(
 \bar{u}_{R(1,I)} \gamma_\mu  u_{R(1,I)} + \bar{d}_{R(1,I)} \gamma_\mu  d_{R(1,I)}
 \right)
+\frac{2}{3}  \bar{u}_{R(2,I)} \gamma_\mu  u_{R(2,I)} 
-\frac{1}{3} \bar{d}_{R(2,I)} \gamma_\mu  d_{R(2,I)}
\right]
 A_2^{3\, \mu}. \nonumber
\end{eqnarray}
Here $g_i$ are the gauge couplings at the $i$-th site. Because of the 
unitarity of the rotation matrix $V_{uL,\,uR,\,dL,\,dR}$, the form of 
the above interaction terms are unchanged with the change of base 
of fermion fields in Eqs.~(\ref{eq:base1})$\sim$(\ref{eq:base2}). 
Thus, there is no FCNC term in the Lagrangian. 

\vspace{5mm}
\subsection{Fermion couplings to charged gauge/NG bosons}

The left-handed fermion couplings to 
charged gauge bosons are expressed as 
\begin{eqnarray}
&& \sum_{I=1}^{3}\left[
\frac{g_0}{\sqrt{2}}  A_0^{+\, \mu}
 \bar{u}_{L(0,I)} \gamma_\mu  d_{L(0,I)}
+  \frac{g_1}{\sqrt{2}}  A_1^{+\, \mu}
 \bar{u}_{L(1,I)}  \gamma_\mu d_{L(1,I)}
 \right]
\ \ +  h.c. ,
\\
&=&
\sum_{I',J'=1}^{3}\left[
\frac{g_0}{\sqrt{2}}  A_0^{+\, \mu}
 \bar{u}_{L(0,I')}' \gamma_\mu  d_{L(0,J')}'
+  \frac{g_1}{\sqrt{2}}  A_1^{+\, \mu}
 \bar{u}_{L(1,I')}' \gamma_\mu  d_{L(1,J')}'
\right]
\left(V_{uL} V_{dL}^\dagger \right)^{I'J'}  + h.c. ,
\\
&=&
\sum_{I',J'=1}^{3}\left[
\frac{g_0}{\sqrt{2}}  A_0^{+\, \mu}
 \bar{u}_{L(0,I')}' \gamma_\mu  d_{L(0,J')}'
+  \frac{g_1}{\sqrt{2}}  A_1^{+\, \mu}
 \bar{u}_{L(1,I')}' \gamma_\mu  d_{L(1,J')}'
\right]
\left(V^{(0)} \right)^{I'J'}  + h.c. ,
\label{eq:charge_int}
\end{eqnarray}
where  in the last equality, we introduced shorthand notation for 
the three-by-three Unitary matrix as
\begin{equation}
\left(V_{uL} V_{dL}^\dagger \right)^{I'J'}\equiv (V^{(0)})^{I'J'}  
\equiv
\left(
\begin{array}{ccc}
  V^{(0)}_{ud} &    V^{(0)}_{us} &    V^{(0)}_{ub} \\
  V^{(0)}_{cd} &    V^{(0)}_{cs} &    V^{(0)}_{cb} \\
  V^{(0)}_{td} &    V^{(0)}_{ts} &    V^{(0)}_{tb}
\end{array}
\right).
\end{equation}
Expression in Eq.~(\ref{eq:charge_int}) 
can be further rewritten in terms of mass eigenstates
using Eqs.~(\ref{eq:Wtrans}) and
(\ref{eq:ftrans}). We show  
the resulting expressions for left-handed fermion couplings to $W$ and 
$W'$ bosons in Appendix \ref{app:Lgauge}

The right-handed fermion couplings to 
charged gauge bosons are expressed as 
\begin{eqnarray}
&& \sum_{I=1}^{3}\left[
\frac{g_1}{\sqrt{2}}  A_1^{+\, \mu}
 \bar{u}_{R(1,I)}  \gamma_\mu d_{R(1,I)}
 \right]
\ \ +  h.c. 
\\
&=&
\sum_{I',J'=1}^{3}\left[
\frac{g_1}{\sqrt{2}}  A_1^{+\, \mu}
 \bar{u}_{R(1,I')}' \gamma_\mu  d_{R(1,J')}'
\right]
\left(V_{uL} V_{dL}^\dagger \right)^{I'J'}  + h.c. 
\\
&=&
\sum_{I',J'=1}^{3}\left[
\frac{g_1}{\sqrt{2}}  A_1^{+\, \mu}
 \bar{u}_{R(1,I')}' \gamma_\mu  d_{R(1,J')}'
\right]
\left(V^{(0)} \right)^{I'J'}  + h.c. 
\end{eqnarray}
Expressions for mass-eigenstate right-handed fermion couplings to $W$ and 
$W'$ bosons are shown in Appendix \ref{app:Rgauge}

Lagrangian for fermion couplings to 
the NG bosons can be written as:

\begin{eqnarray}
&& - 2 i \sum_{I,J=1}^{3}
\left(
\begin{array}{cc}
\bar{u}_{L(0,I)} & 
\bar{d}_{L(0,I)} 
\end{array}
\right) 
         \left( 
           \begin{array}{c|c}
               & \lambda\, \delta_{IJ}\, \frac{\pi_1^+}{\sqrt{2}}  \\ \hline
              \lambda\, \delta_{IJ}\, \frac{\pi_1^-}{\sqrt{2}} &   \\ 
           \end{array}
         \right)
         \left( 
           \begin{array}{c}
              u_{R(1,J)}\\
              d_{R(1,J)}
         \end{array}
         \right) \nonumber \\
&-& 2 i \sum_{I,J=1}^{3}
\left(
\begin{array}{cc}
\bar{u}_{L(1,I)} & 
\bar{d}_{L(1,I)} 
\end{array}
\right) 
         \left( 
           \begin{array}{c|c}
               & (\Lambda_d')_{IJ}\, \frac{\pi_2^+}{\sqrt{2}}  \\ \hline
              (\Lambda_u')_{IJ}\, \frac{\pi_2^-}{\sqrt{2}} &   \\ 
           \end{array}
         \right)
         \left( 
           \begin{array}{c}
              u_{R(2,J)}\\
              d_{R(2,J)}
         \end{array}
         \right) \nonumber \\
&+& h.c.
\end{eqnarray}
Forms of couplings in terms of mass eigenmodes are summarized in Appendix 
\ref{app:NG}.

\vspace{1cm}
\section{$b \rightarrow s \gamma$ in the Three Site Higgsless Model}
\label{sec:bsgamma}

In the previous section, we introduced a generation structure into the 
three site Higgsless model, and derived all the couplings among 
light/heavy fermions and gauge/NG bosons. Thus, it is ready 
to calculate the amplitude of any flavor changing process.
Here, we calculate the branching ratio of the $b \rightarrow s \gamma$ 
process in the three site Higgsless model, and compare it to  
experimental results.

\vspace{5mm}
\subsection{Contributions to $b \rightarrow s \gamma$ at $M_W$ scale}
Constraint on $\varepsilon_{tR}$ 
from the $b \rightarrow s \gamma$ measurement 
due to the right-handed light fermion coupling to the $W$ boson 
was discussed in Ref.~\cite{SekharChivukula:2006cg} 
taking advantage of the analysis done in Ref.~\cite{Larios:1999au}. 
However, the contribution from the flavor mixing interaction between
the light fermion and the heavy partner has never been studied
before. 
Here, we calculate all the leading contribution to the 
one-loop $b \rightarrow s \gamma$ amplitude including diagrams 
which involve the propagator(s) of heavy states. In the following 
presentation, we use the same notation as in Ref.~\cite{Chetyrkin:1996vx}

\begin{figure}[t] 
 \centering
  \includegraphics[width=0.32\textwidth]{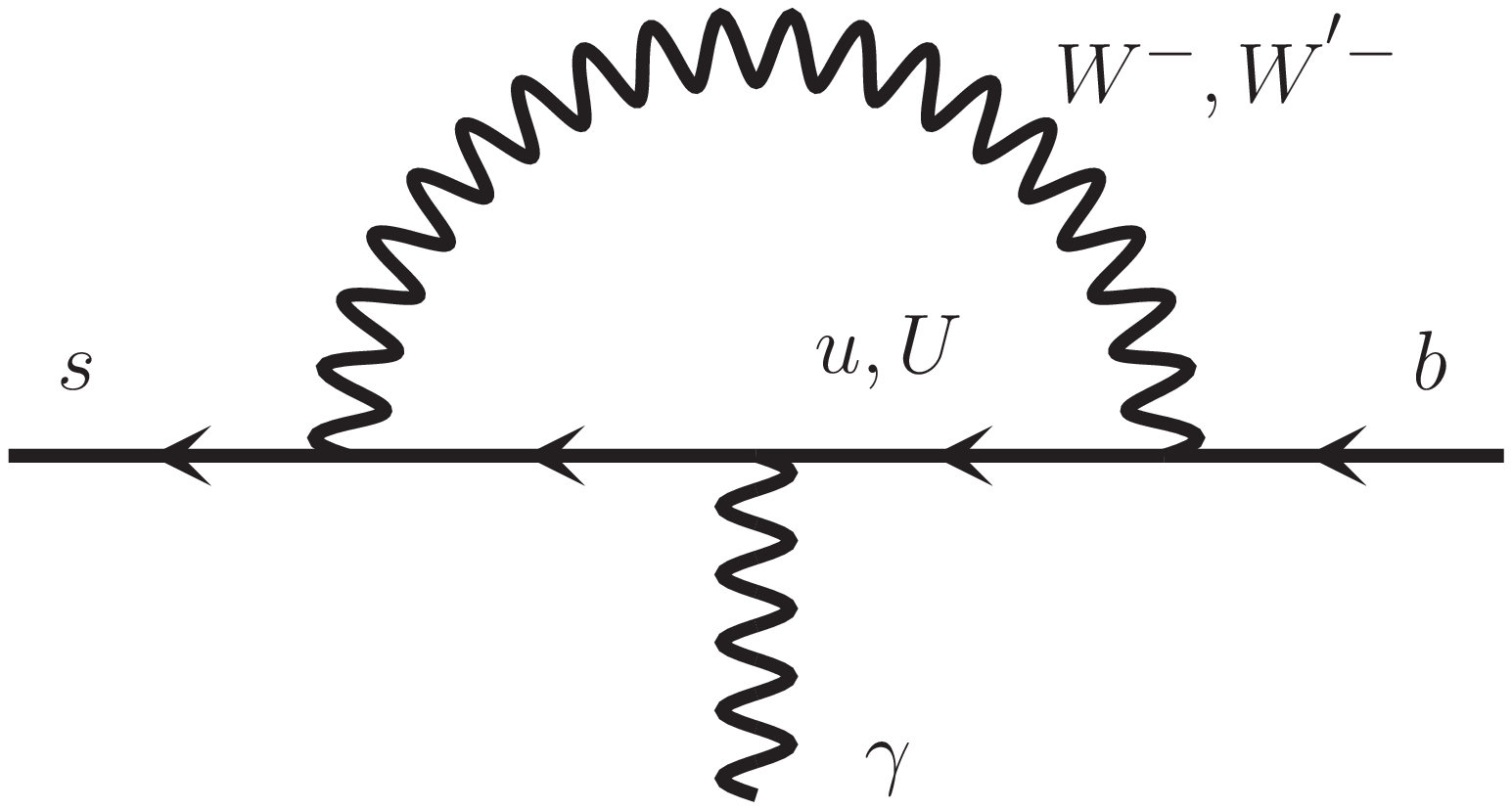}
  \includegraphics[width=0.32\textwidth]{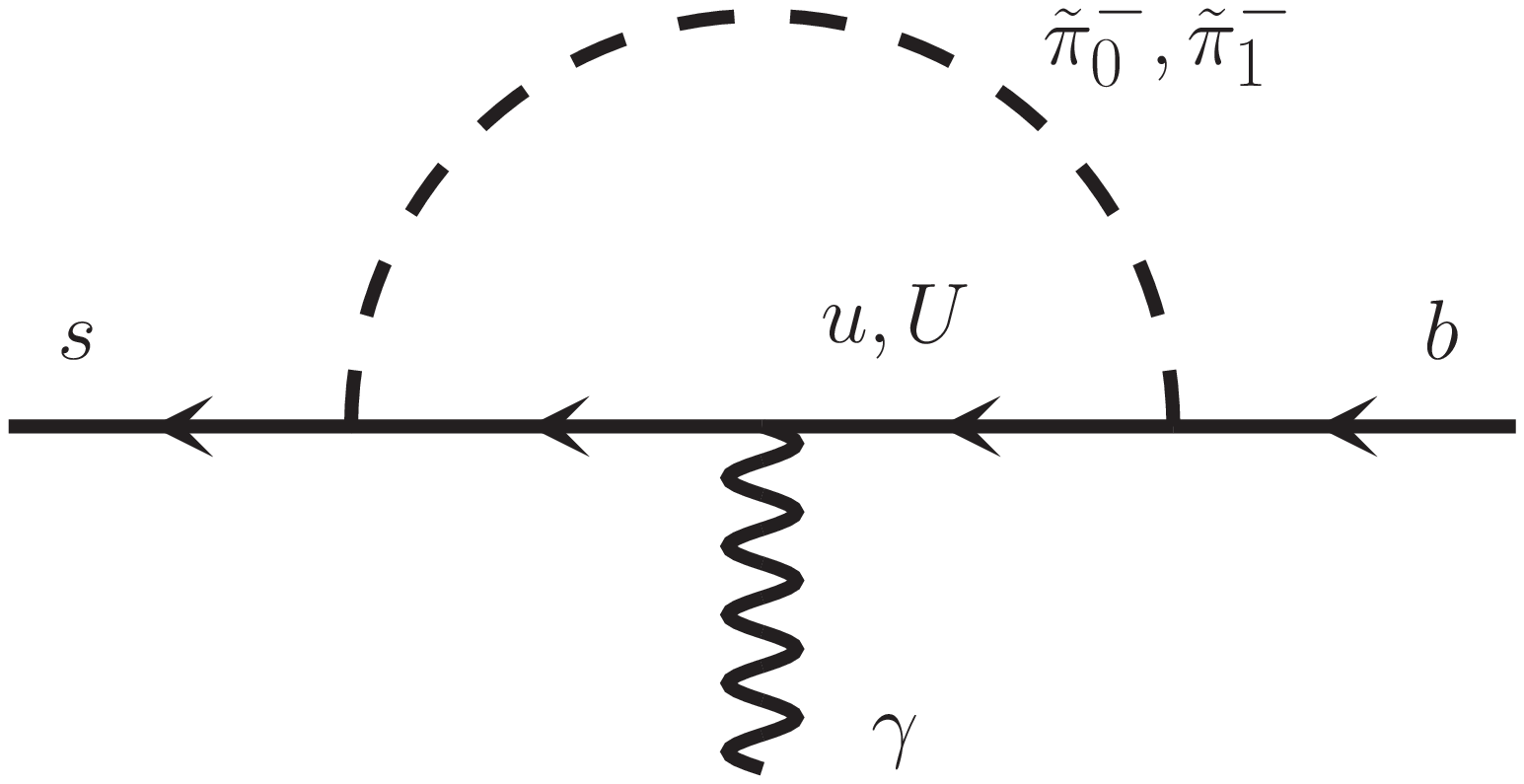}
  \includegraphics[width=0.32\textwidth]{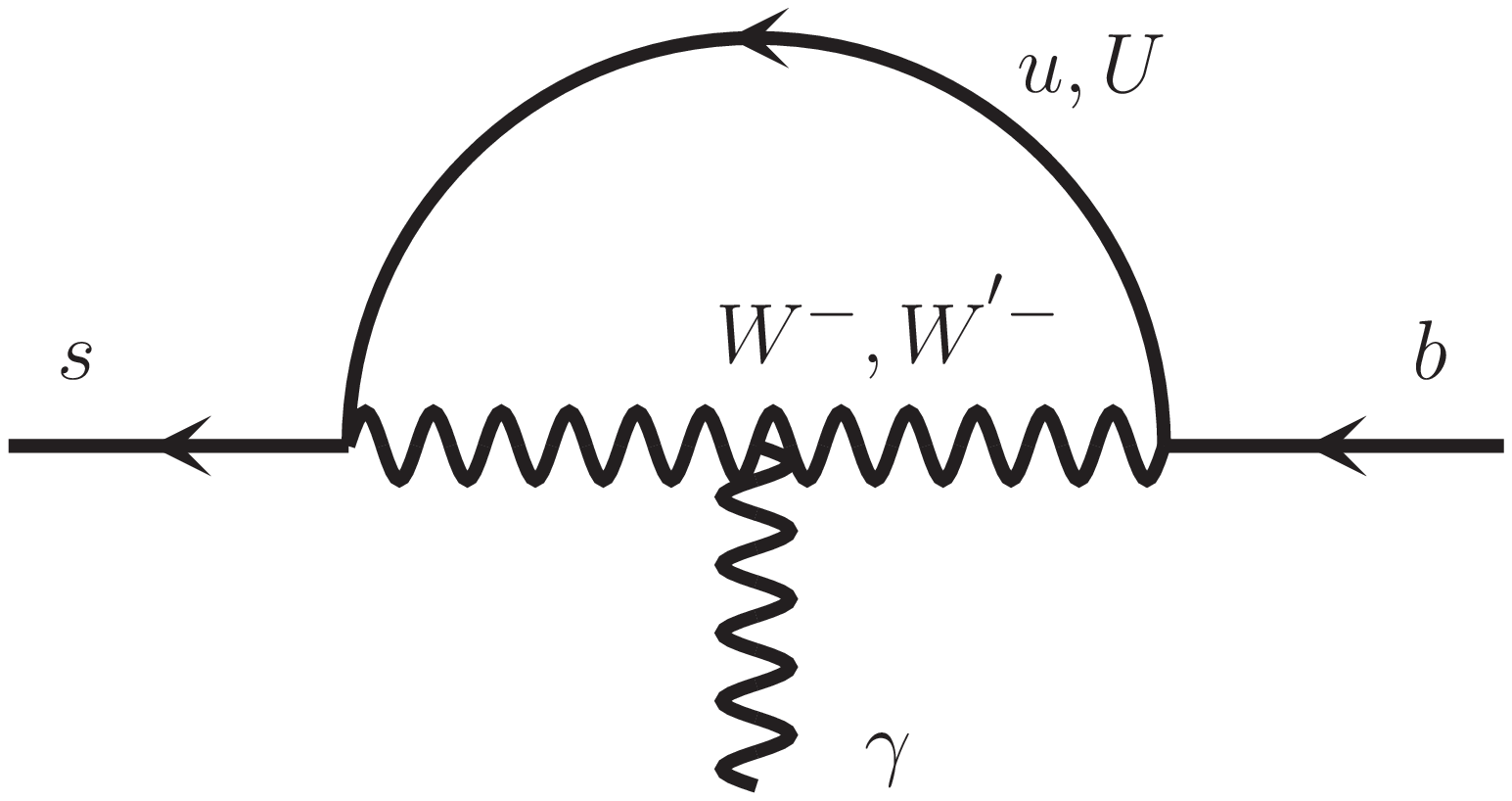}\vspace{5mm}
  \includegraphics[width=0.32\textwidth]{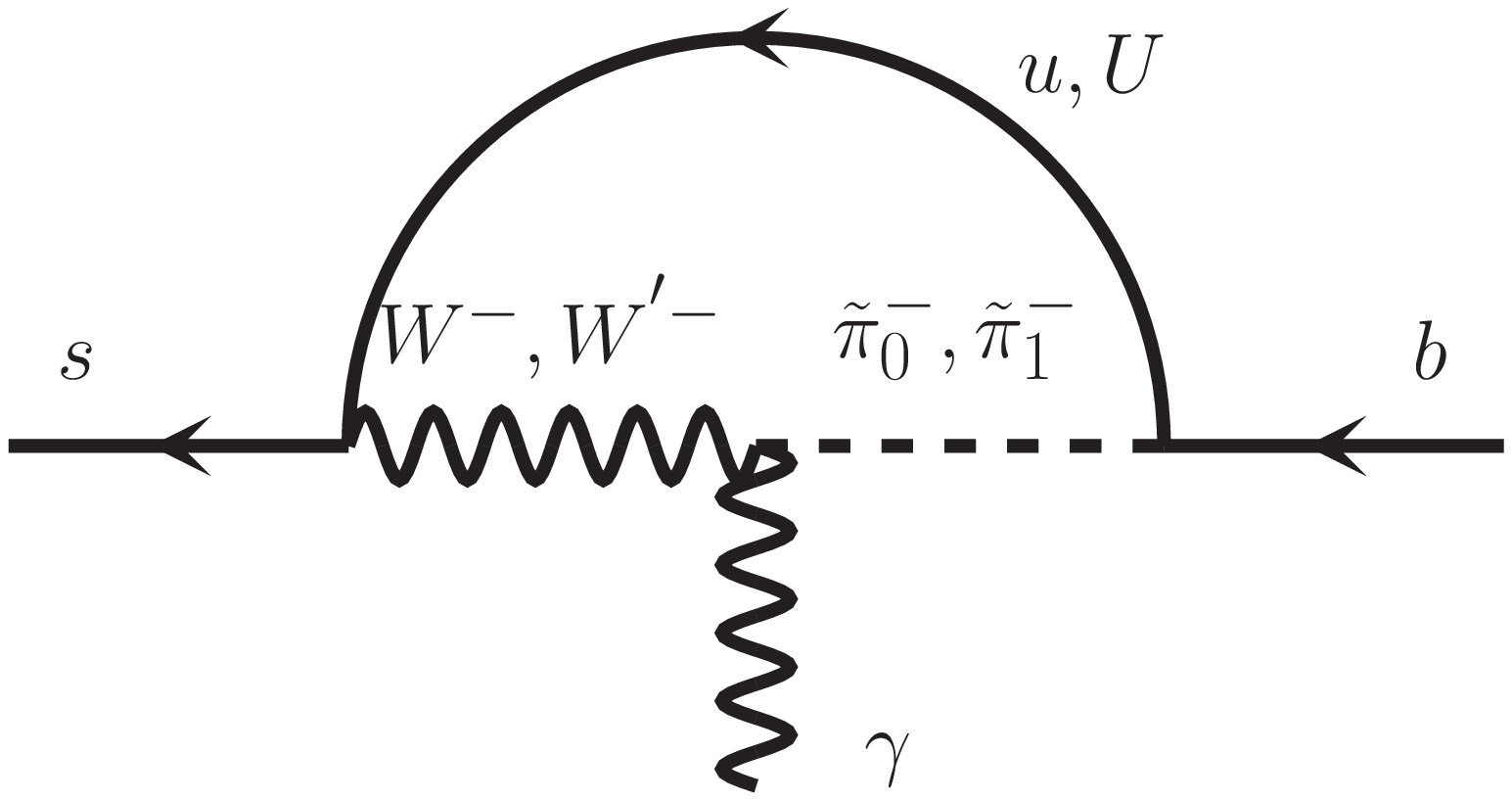}
  \includegraphics[width=0.32\textwidth]{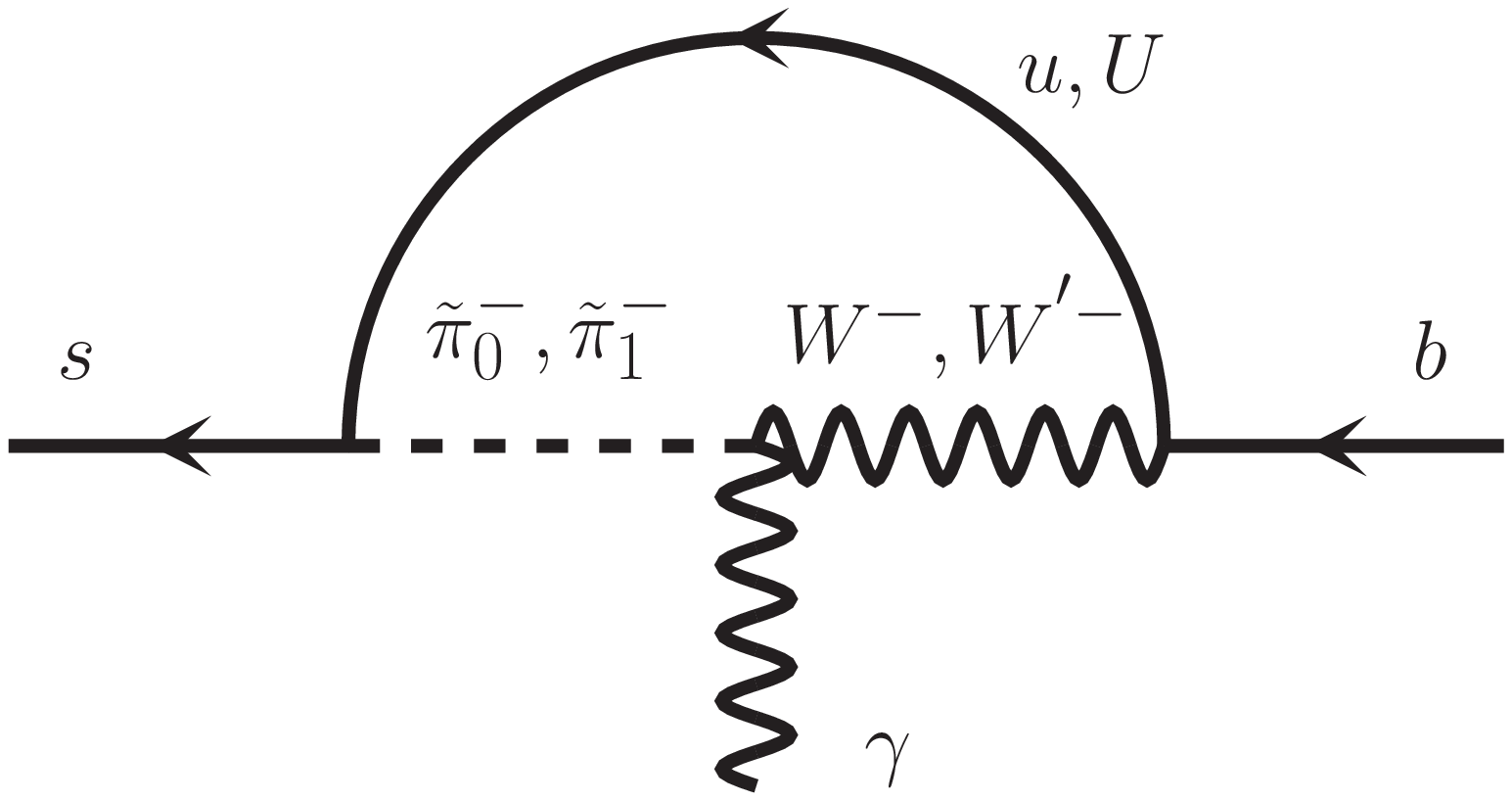}
  \includegraphics[width=0.32\textwidth]{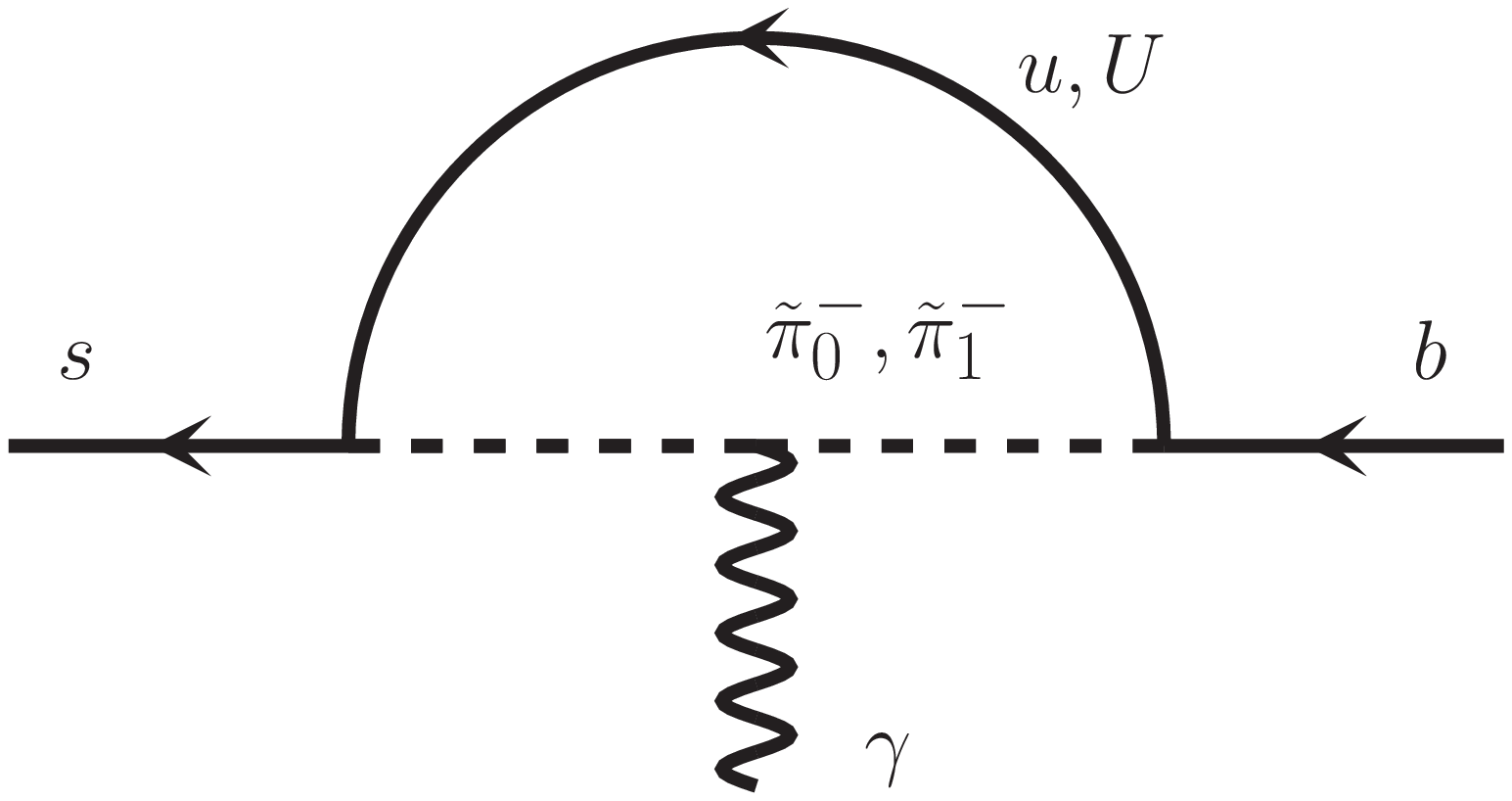}
 \caption{One-loop diagrams contributing to the induced $b \bar{s} \gamma$ vertex. 
 In the figure, $u, U$ represent all the up-type quarks, namely $u, c, t$ and their 
 heavy partners. Upper-left and upper-middle diagrams with the photon replaced 
 by the gluon give the induced $b \bar{s}$-gluon vertex. }
 \label{fig:one-loop}
\end{figure}

One-loop diagrams which contribute to  $C_{7}^{(0)}(M_W)$ are shown in 
Fig.~\ref{fig:one-loop}. 
Diagrams which contribute to  $C_{8}^{(0)}(M_W)$ can be obtained by replacing the 
photon by the gluon in the upper-left and upper-middle diagrams in Fig.~\ref{fig:one-loop}. 
First, let us discuss the general features of one-loop
  amplitudes. The cancellations of ultraviolet (UV) divergences
  take place after summing all the diagrams in
  Fig.~\ref{fig:one-loop} and contributions from self-energy 
  diagrams of the external fermions. 
  If we separate the UV divergences, the GIM
  mechanism tells us that each diagram can give nonvanishing contributions 
  due to the difference in the masses and couplings of the particles in the
  loop. Therefore, in principle heavy partner fields can give
  sizable corrections to $b\rightarrow \gamma$ amplitude. This can
  be understood as the non-decoupling effect since the masses of the
  heavy partner fields are proportial to yukawa coupling. 

Now, let us look into the contributions from the light
  and heavy particles more in detail. 
Among diagrams which involve only light particles in their loops, those which 
also exist in the SM take the same values as in the case of the SM at the leading 
order in the expansion of $x^2$. (See Refs. \cite{Inami:1980fz,Chetyrkin:1996vx} 
for the SM expressions.) 
This is because leading expressions of couplings 
among SM particles take the SM forms due to the ideal fermion delocalization.
There are also diagrams which involve only light particles, but 
do not exist in the SM. An example is the upper-left diagram in Fig.~\ref{fig:one-loop} 
with right-handed fermions couple to $W$ boson. 
The leading contribution to $C_7^{(0)}(M_W)$ 
from it is expressed as
\begin{equation}
Q_u\ \frac{\varepsilon_{tR}^2}{2(1+\varepsilon_{tR}^2)}
\frac{1+2 x_t\log x_t - x_t^2}{(x_t-1)^3}, 
\end{equation}
where, $Q_u (=+\frac{2}{3})$ is the electric charge of up-type fermions, and 
$x_t$ is defined as $x_t \equiv \frac{m_t^2}{m_W^2}$.
Here, we neglected the small CKM matrix element $V^{(0)}_{ub}$ 
and used the unitarity relation $V^{(0)\ast}_{cb}V^{(0)}_{cs} = - V^{(0)\ast}_{tb}V^{(0)}_{ts}$.
This contribution vanishes in the limit of $\varepsilon_{tR}\rightarrow 0$ because 
the right-handed fermion couplings to the $W$ boson vanish in that limit.

As for diagrams which involve at least one heavy particle in the loop, 
in general, propagators of heavy particles give suppression factor of $x^2$ 
because of the mass hierarchy between light and heavy particles: 
$\frac{M_W^2}{M_{W'}^2} = O(x^2), 
\frac{m_f^2}{m_F^2} = O(\varepsilon_L^2) = O(x^2)$.
However, as can be seen in the expressions summarized in 
Appendix \ref{app:Fcouplings}, 
some of couplings which involve heavy particle(s) have enhancement 
factor of $1/x$. Because of this, there are diagrams with heavy particles which 
contribute to $C_{7, 8}^{(0)}(M_W)$ at the same order as contributions 
from diagrams with only light particles. 
As an example, we show the leading expression of the contribution to $C_7^{(0)}(M_W)$ 
from upper-middle diagram 
in Fig.~\ref{fig:one-loop} with both fermion and gauge bosons being heavy 
states:
\begin{equation}
Q_u \left(
\frac{z_t(3-4z_t+z_t^2+2\log z_t)}{16(z_t-1)^3}
-  \frac{z_c(3-4z_c+z_c^2+2\log z_c)}{16(z_c-1)^3}
\right).
\label{eq:heavy}
\end{equation}
Here, with neglecting $\varepsilon_{cR}^2 (\ll 1)$, $z_t$ and $z_c$ are 
expressed as 
$z_t \equiv \frac{m_T^2}{m_{W'}^2} = \frac{m_t^2}{m_W^2}
\frac{(1+\varepsilon_{tR}^2)^2}{2\varepsilon_{tR}^2}[1+O(x^2)]$, 
$z_c \equiv \frac{m_C^2}{m_{W'}^2} = \frac{m_t^2}{m_W^2}
\frac{(1+\varepsilon_{tR}^2)}{2\varepsilon_{tR}^2}[1+O(x^2)]$. 
The first and second terms in the above expression represent the 
contributions from heavy partners of the top and charm quarks. 
We should note that Eq.~(\ref{eq:heavy}) vanishes in the limit of 
$\varepsilon_{tR}\rightarrow 0$, in which masses 
of heavy-partner fermions become infinity. 
This is not because of the decoupling of the heavy fermions 
(each term in Eq.~(\ref{eq:heavy}) does not vanish), but because
GIM-like mechanism works in the three site Higgsless model. 
(Note that $\lim\limits_{\varepsilon_{tR} \to 0} (z_t/z_c) = 1$.)

\vspace{5mm}
\subsection{Contributions to $b \rightarrow s \gamma$
at low energy scale}
The leading-order $b \rightarrow s \gamma$ matrix element of the effective 
Hamiltonian is proportional to the leading-order term in the effective coefficient $C_7^{\rm eff}$. 
Its expression at the scale $\mu_b$ (which will be taken to be $m_b$ in our numerical 
estimation) can be written as~\cite{Chetyrkin:1996vx}:
\begin{equation}
C_7^{(0) {\rm eff}}(\mu_b) 
= \eta^{\frac{16}{23}} C_7^{(0)}(M_W) 
+ \frac{8}{3}
\left(
\eta^{\frac{14}{23}}  - \eta^{\frac{16}{23}} 
\right)
C_8^{(0)}(M_W)  
+
\sum_{i=1}^{8} h_i \eta^{a_i}, 
\label{eq:effective}
\end{equation}
where $\eta=\alpha_s(M_W)/\alpha_s(\mu_b)$ with $\alpha_s(\mu)$ being the strong 
coupling constant at the scale of $\mu$, and $C_{7}^{(0)}(M_W)$, $C_{8}^{(0)}(M_W)$ 
are leading order coefficients of the magnetic and chromomagnetic operators 
at the scale of $M_W$. The third term in Eq.(\ref{eq:effective}) comes from the 
mixing with four-fermion operators with their coefficients at $M_W$ scale 
$C^{(0)}_{1,3,4,5,6}(M_W)=0$ and 
$C^{(0)}_{2}(M_W)=1$\footnote{Strictly speaking, there are induced contributions to 
coefficients of four-fermion operator from the heavy-gauge-boson exchange. 
However, those are negligible because $W'$ and $Z'$ have heavy masses  
and couplings of them to light fermions are highly suppressed because of the 
ideal fermion delocalization.}, 
and the values of $h_i$ and $a_i$ being 
$a_i=\left(\frac{14}{23}, \frac{16}{23}, \frac{6}{23}, -\frac{12}{23},
0.4086, -0.4230, -0.8994, 0.1456 \right)$,
$h_i=\left(\frac{626126}{272277}, -\frac{56281}{51730}, -\frac{3}{7},
-\frac{1}{14}, -0.6494, -0.0380, -0.0186, -0.0057 \right)$.

\vspace{5mm}
\subsection{Numerical analysis}

Before showing the result of one-loop calculations, let us recall the model parameters 
in the current study. In the gauge sector of the model, there are four free paramters: 
$f$, $g_0 (\equiv g)$, $g_1 (\equiv \tilde{g})$, and $g_2 (\equiv g')$. Three of these are 
chosen to reproduce three electroweak observables (say, $m_W$, $e$ and $\sin \theta_W$), 
and remaining one is used to fix the scale of heavy gauge boson mass 
($m_{W'}$, for example).  
As for fermion sector, there are $3 \times 3$ Unitary mixing matrix $V^{(0)}$, 
the scale of heavy fermion $M$, flavor universal parameter $\varepsilon_L$, 
and flavor dependent parameter $\varepsilon_{fR}$. At the leading approximation, 
values of each element of $V^{(0)}$ can be taken to be same as those of the SM 
Cabibbo-Kobayashi-Maskawa (CKM) matrix. Value of $\varepsilon_L$ is fixed 
from the requirement of the ideal fermion delocalization 
once the value of $x (\equiv g/\tilde{g})$ is chosen. (See Eq.~(\ref{eq:ideal}).) 
Among $\varepsilon_{fR}$ for different flavors, $\varepsilon_{tR}$ is the one 
which is relevant to $b \rightarrow s \gamma$ amplitude, and we take this 
as an input parameter in the following calculations. As for $M$, it has to be 
chosen in a way that the top quark mass is correctly reproduced through 
Eq.~(\ref{eq:fermion_mass}). Thus, for a given value of $\varepsilon_L$ 
(which is fixed by the a chosen value of $M_W/M_{W'}$), there is 
one-to-one correspondence between $M$ and $\varepsilon_{tR}$.

\begin{figure}[t] 
 \centering
 \includegraphics[width=0.46\textwidth]{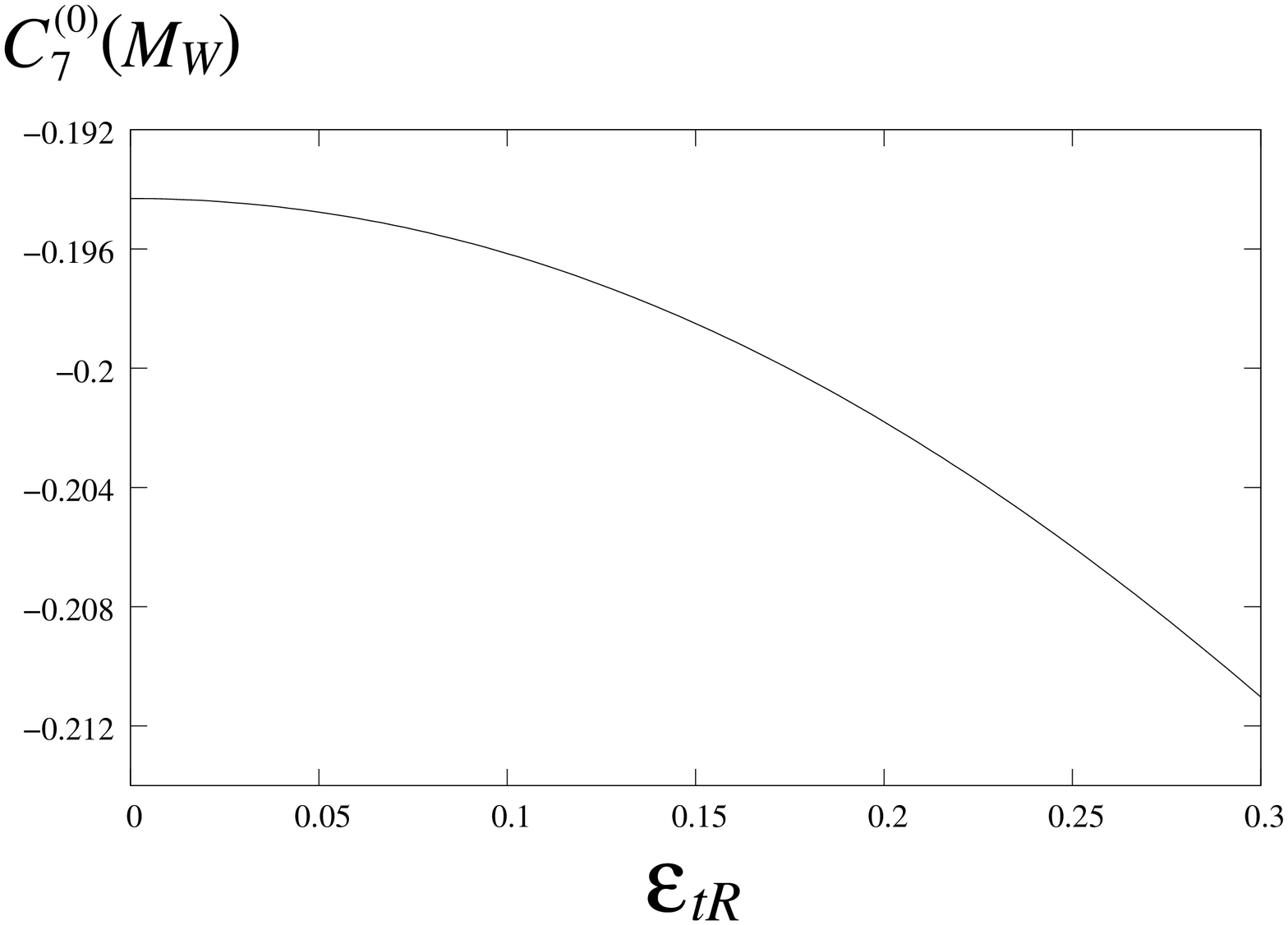}\ \ \ \ 
 \includegraphics[width=0.455\textwidth]{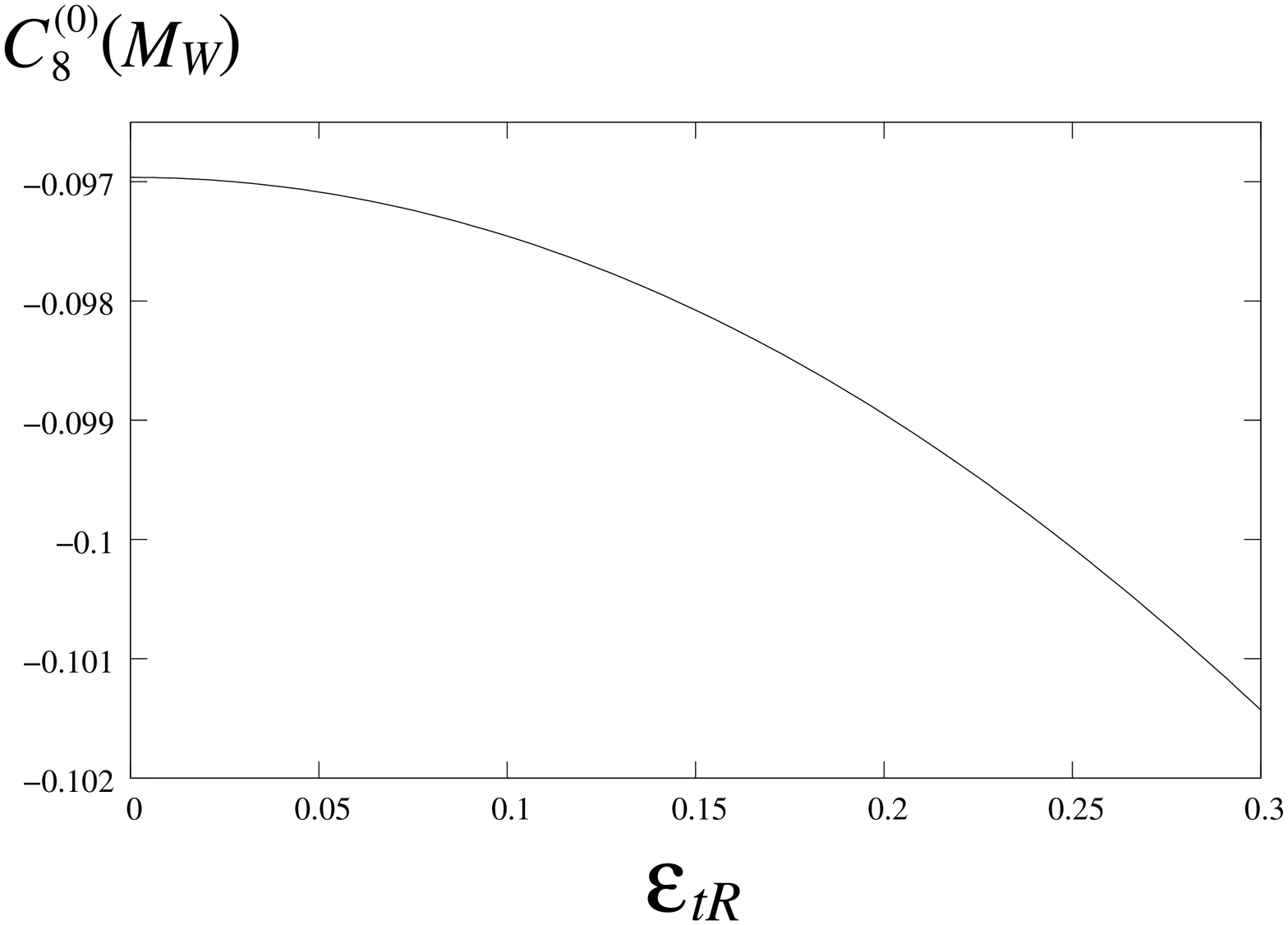}
 \caption{Values of $C_7^{(0)}(M_W)$ (left panel) and $C_8^{(0)}(M_W)$ 
 (right panel) as functions of 
 $\varepsilon_{tR}$.}
 \label{fig:C7C8MW}
\end{figure}
Let us now turn to the numerical results. 
In Fig.~\ref{fig:C7C8MW}, we show values of $C_7^{(0)}(M_W)$ 
and $C_8^{(0)}(M_W)$ as functions of $\varepsilon_{tR}$. 
In the numerical evaluations, we keep only the leading contributions 
in the expansion of $x^2$ and $\varepsilon_L^2$. 
Values of them could take 
$0.009 < \varepsilon_L^2 (\simeq x^2/2) < 0.09$ depending 
on the choise of $W'$ mass within 380 GeV $<$ $M_{W'}$ $<$ 1.2 TeV. 
Here, the lower bound on $M_{W'}$ comes from the multi-gauge-boson 
measurement, while the upper bound is set requiring perturbative unitarity.
Since the leading contribution to the $b\rightarrow s \gamma$ amplitude 
is $O(x^0)$ (or $O(\varepsilon^0$)), 
the choice of values of $x^2$ ($\varepsilon_L^2$) does not change 
our leading order estimation. The choice of $x^2$ ($\varepsilon_L^2$) 
rather determine the magnitude of uncertainty 
due to the truncation of higher order terms, which are naively expected 
to be 1 $\sim$ 10\% for an allowed value of $x^2$ ($\varepsilon_L^2$) quoted 
above. The small CKM matrix element 
$V^{(0)}_{ub}$ as well as the s-quark mass are neglected in the present 
analysis. As input values for the $W$ and top quark masses, we took 
$M_W \simeq 80.4$ GeV and $m_t = m_t^{\overline{MS}}(M_W) \simeq 172.9$ 
GeV. As we mentioned at the end of
Section~\ref{overview},  we show the result for $0 \le
\varepsilon_{tR} \le 0.3$  
since a value larger than $\varepsilon_{tR} \sim 0.3$ would cause large 
isospin violation which is inconsistent with the precision electroweak 
measurement~\cite{SekharChivukula:2006cg,Abe:2008hb}. 
The values of $C_{7, 8}^{(0)}(M_W)$ at $\varepsilon_{tR}=0$ coincide 
with those in the case of the SM. Thus, from the plots in 
Fig.~\ref{fig:C7C8MW}, we see that magnitude of coefficient $C_{7}^{(0)}(M_W)$ 
$\big( C_{8}^{(0)}(M_W) \big)$ in the three site Higgsless model 
would increase by about 8.5\% (4.5\%)
from that in the SM.

From Eq.~(\ref{eq:effective}), with these results of $C_7^{(0)}(M_W)$ 
and $C_8^{(0)}(M_W)$, we obtain the value of $C_7^{(0) {\rm eff}}(\mu_b)$.
\begin{figure}[t] 
 \centering
 \includegraphics[width=0.6\textwidth]{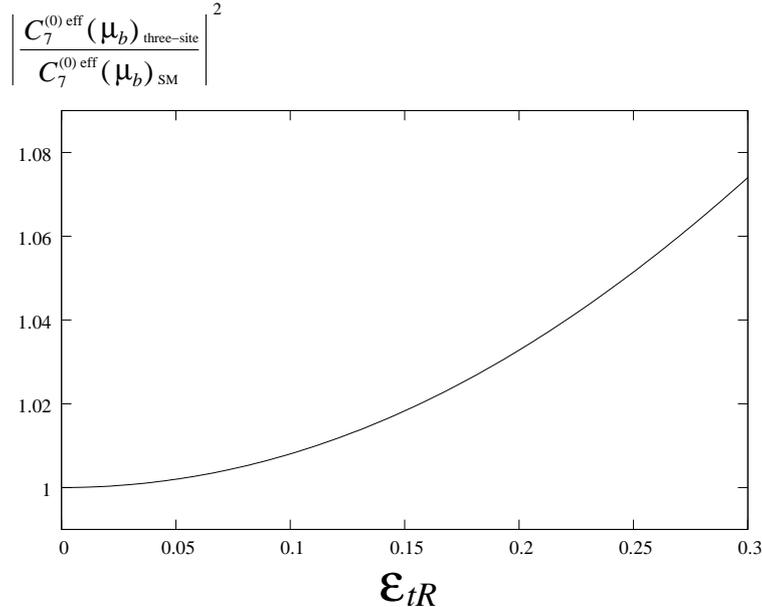} 
 \caption{The ratio of the value of $\vert C_7^{(0) {\rm eff}}(\mu_b) \vert^2$ 
 in the three site Higgsless model to that in the SM as a function 
 of $\varepsilon_{tR}$.}
 \label{fig:ratio}
\end{figure}
In Fig.~\ref{fig:ratio}, we plot the ratio of the value of 
$\vert C_7^{(0) {\rm eff}}(\mu_b) \vert^2$ in the three site Higgsless model 
to that in the SM. Here, we used $\mu_b = m_b \simeq 4.2$ GeV, 
in which case the value of $\eta$ is estimated as $\eta \simeq 0.54$.
From the figure, we see that $\vert C_7^{(0) {\rm eff}}(\mu_b) \vert^2$ in the 
three site Higgsless model would increases by about 7.4\%
from that in the SM.

\begin{figure}[t] 
 \centering
 \includegraphics[width=0.7\textwidth]{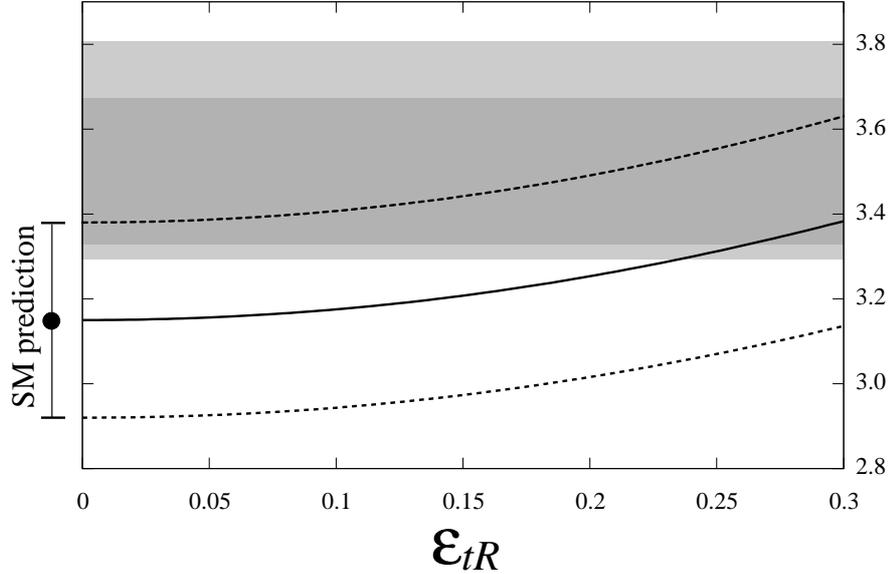} 
 \caption{$B \rightarrow  X_s \gamma$ branching ratio in the 
 three site Higgsless model (middle curve) as a function of 
 $\varepsilon_{tR}$. Top and bottom dashed curves indicate 
 upper and lower edges of the uncertainty. 
 Light and dark shaded areas indicate 
 the experimental data 
 ${\bf {\cal B}}(B\rightarrow X_s \gamma,
 E_{\gamma}>1.6 \mbox{GeV})_{\rm exp}  
 = (3.55 \pm 0.24 \pm 0.09) \times 10^{-4}$, and 
 ${\bf {\cal B}}(B\rightarrow X_s \gamma,
 E_{\gamma}>1.6 \mbox{GeV})_{\rm exp} 
 = (3.50 \pm 0.14 \pm 0.10) \times 10^{-4}$, 
 reported respectively in Ref.~\cite{Barberio:2008fa} and 
 Ref.~\cite{Artuso:2009jw}. The point with an error bar 
 placed on the left of the plot indicates the SM 
 prediction~\cite{Misiak:2006zs}.
 }
 \label{fig:Br}
\end{figure}

There are several steps to evaluate the value of the branching ratio 
of $B \rightarrow X_s \gamma$ from the value of 
$\vert C_7^{(0) {\rm eff}}(\mu_b) \vert^2$, and careful studies are 
needed to estimate its uncertainty. Here, to evaluate the central value 
of the $B \rightarrow X_s \gamma$ branching ratio and its uncertainty 
in the three site Higgsless model, we simply multiply the enhancement 
factor plotted in Fig.~\ref{fig:ratio} to those values of the SM prediction.
As an input SM value, we take the value 
reported in Ref.~\cite{Misiak:2006zs}, namely 
${\bf {\cal B}}(B\rightarrow X_s \gamma)_{\rm SM} = (3.15 \pm 0.23) 
\times 10^{-4}$
\footnote{We exploit the next-to-leading order result
for the standard model as an overall normalization, while the
relative enhancement factor due to the new physics effect is evaluated only
at the leading order. Therefore, in our estimate, the next-to-leading
order correction to the new physics effect is neglected. 
Since the new physics effect is less than 10\%, this approximation
should be accpetable for the purpose of our current study.}.
In Fig.~\ref{fig:Br}, we show the resulting value of 
${\bf {\cal B}}(B\rightarrow X_s \gamma)$ in the three site Higgsless model 
as a function of $\varepsilon_{tR}$ (middle solid curve). Upper and lower 
dashed curves indicate upper and lower edges of the
uncertainty. 
 Light and dark shaded areas indicate 
 the experimental data,
 ${\bf {\cal B}}(B\rightarrow X_s \gamma,
 E_{\gamma}>1.6 \mbox{GeV} )_{\rm exp}  
 = (3.55 \pm 0.24 \pm 0.09) \times 10^{-4}$, and 
 ${\bf {\cal B}}(B\rightarrow X_s \gamma,
 E_{\gamma}>1.6 \mbox{GeV} )_{\rm exp}  
 = (3.50 \pm 0.14 \pm 0.10) \times 10^{-4}$, 
 reported respectively in Ref.~\cite{Barberio:2008fa} and 
 Ref.~\cite{Artuso:2009jw}.
Note that the value at $\varepsilon_{tR}=0$ coincides 
with that in the case of the SM prediction, which is indicated 
on the left of the plot in the figure. 
The result shows that the central value of the $B \rightarrow X_s \gamma$ 
branching ratio in the three site Higgsless model takes closer value 
to its experimental central value as one takes the larger value of 
$\varepsilon_{tR}$.

\vspace{5mm}
\subsection{Discussions}
The three site Higgsless model can be considered as a low-energy 
effective model of any theories which have a vector resonance below 
or around a TeV scale. (Extra-dimensional models and Technicolor-type 
models are possible candidates.)
Such high-energy theory  would produce  higher-dimensional operators 
at the cutoff scale of the three site Higgsless model, and  
some of those contribute to the $b \rightarrow s \gamma$ amplitude.
Once we specify the theory above the cutoff scale, we can, in principle, 
calculate coefficients of those higher-dimensional operators and 
estimate contributions to the $b\rightarrow s \gamma$ process. 
(See, for example, Ref.~\cite{Randall:1993ua}~, which discussed 
contributions to the $b\rightarrow s \gamma$ amplitude in the 
extended Technicolor models. It was shown that there was at most 
about 10\% addition to the SM value.)
We did not try to include such contributions from physics 
above the cutoff scale in our calculation in the previous subsection. 
Instead, we estimated the model-independent contributions 
to the $b \rightarrow s \gamma$ amplitude which come from physics below the 
cutoff scale. Thus, we have to be aware that the contribution from the 
high-energy theory should be added to the value obtained in the 
previous subsection. Having said that, it is interesting to point out 
that our result 
in the previous subsection indicates that even if there was 10\% 
modification of the $b \rightarrow s \gamma$ amplitude due to  
contributions from the high-energy theory, 
it is still consistent with experiment at $1 \sigma$ level.
It is remarkable that in the wide range of the parameter
space the three site Higgsless model is compatible with the
experimental data for $B\rightarrow X_s \gamma$.

\vspace{5mm}
\section{Conclusions}
\label{sec:conclusions}

In this paper, we have explicitly shown a minimal way of introducing flavor  
structure into the three site Higgsless model, 
and derived explicit forms of fermion couplings to gauge/NG bosons 
with including flavor mixings. 
Generation structure was 
introduced in a way that it respects the flavor symmetry, thus there is 
no FCNC at tree level. Because 
of the flavor mixing, FCNC's appear at one-loop level. As an example of 
such FCNC processes, we have calculated the contribution to the 
$b \rightarrow s \gamma$ process in the three site Higgsless model. 
We have checked that all divergences cancels 
because of the existence of GIM-like mechanism, providing finite 
$b \rightarrow s \gamma$ amplitude at one-loop level.  
We have shown that heavy particles do not completely decouple 
in the heavy-mass limit because some couplings which involve 
heavy particle(s) have enhancement factors. One-loop level 
$b \rightarrow s \gamma$ amplitude was calculated by considering  
all possible combination of particles in the loop, and the leading expression 
in the expansion of $x^2 (\simeq 4M_W^2/M_{W'}^2)$ was derived.
The result showed that the central value of the $B \rightarrow X_s \gamma$ 
branching ratio in the three site Higgsless model takes closer value 
to its experimental central value as one takes the larger value of 
$\varepsilon_{tR}$, and indicated that even if there was, say, 10\% 
modification of the $b \rightarrow s \gamma$ amplitude due to  
contributions from possible high-energy theory, 
it is still consistent with experiment at $1 \sigma$ level.
It will be interesting to see how the constraints in
the $B\rightarrow K^{\ast} l^+ l^-$ decay would confront the Higgsless
model with the experimental data from LHC-b and future super B
factories followed by the theoretical improvments in lattice
QCD. Another important FCNC process which is senstive to the third
generation is the $B_s-\overline{B}_s$ mixing, which has drawn much
attention by the recent measurement of the charge asymmetry at
D0~\cite{Abazov:2010hv}.

\vspace{5mm}
\section*{Acknowledgements}
We thank Ryuichiro Kitano, Tomohiro Abe and Masaharu Tanabashi 
for valuable comments. M.K. also thanks 
R. Sekhar Chivukula, Elizabeth H. Simmons, Hong-Jian He and 
Masaharu Tanabashi for continuing discussions on Higgsless models.
This work is supported by the Grant-in-Aid of the Japanese Ministry 
of Education (Nos. 20105002).

\vspace{1cm}
\appendix

\section{Details of the gauge sector of the three site Higgsless model}
\label{app:gauge}

\subsection{Gauge bosons}
The mass squared matrix for the neutral gauge bosons is given by 
\beq
 {\cal M}_Z^2 = \frac{g_1^2 v^2}{2}
 \left(
 \begin{array}{ccc}
  x^2 & - x & 0  \\
  - x & 2 & - t x \\
  0  & - t x &  t^2 x^2
 \end{array}
 \right) ,
\eeq
where $t \equiv \frac{\sin{\theta}}{\cos{\theta}} \equiv \frac{g_2}{g_0}$, and 
$x \equiv \frac{g_0}{g_1}$. The matrix can be diagonalized 
by the orthogonal rotation, 
\beq
{R_z}^T {\cal M}_Z^2 R_z = \left({\cal M}_Z^2\right)^{\rm diag} 
 = 
\left(
 \begin{array}{ccc}
  M_\gamma^2 (=0)  &   & \\
  & M_{Z}^2 & \\
  & & M_{Z'}^2
 \end{array}
 \right) ,
\eeq
where each column vector in $R_z$ correspond to the eigenvector 
for photon, $Z$, and $Z'$ bosons, respectively:
\beq
R_z
 = 
\left(
 \begin{array}{ccc}
  v_\gamma^0 & v_Z^0 & v_{Z'}^0 \\
  v_\gamma^1 & v_Z^1 & v_{Z'}^1 \\
  v_\gamma^2 & v_Z^2 & v_{Z'}^2 
     \end{array}
 \right).
\eeq
The photon profile, which is needed for the calculation of one-loop diagram contributing 
to $b \rightarrow s \gamma$, is given by 
\beq
\left(v_\gamma^0, v_\gamma^1, v_\gamma^2 \right) 
= 
\left( \frac{e}{g_0}, \frac{e}{g_1}, \frac{e}{g_2} \right), 
\eeq
where the normalization constant $e$, which is identified as the electromagnetic 
charge, is defined as 
\beq
\frac{1}{e^2} = \frac{1}{g_0^2} + \frac{1}{g_1^2} + \frac{1}{g_2^2}.
\eeq

The mass squared matrix for the charged gauge bosons is given by 
\beq
 {\cal M}_W^2 = \frac{g_1^2 v^2}{2}
 \left(
 \begin{array}{cc}
  x^2 & - x   \\
  - x & 2 
 \end{array}
 \right), 
\eeq
which is diagonalized by the orthogonal rotation as
\beq
{R_w}^T {\cal M}_W^2 R_w = \left({\cal M}_W^2\right)^{\rm diag} 
 = 
\left(
 \begin{array}{cc}
 M_{W}^2 & \\
& M_{W'}^2
 \end{array}
 \right) ,
\eeq
where each column vector in $R_w$ correspond to the eigenvector 
for $W$ and $W'$ bosons, respectively:
\beq
R_w
 = 
\left(
 \begin{array}{cc}
  v_W^0 & v_{W'}^0 \\
  v_W^1 & v_{W'}^1 
     \end{array}
 \right).
\eeq
For the case of $x^2 \ll 1$, the mass eigenvalues of $W$ and $W'$ gauge 
bosons can be expanded as 
\beqn
M_W^2 &=& \left(\frac{g_0 v}{2}\right)^2 \left[ 1 - \frac{x^2}{4} + \cdots \right] ,
\label{eq:Wmass}\\
M_{W'}^2 &=& \left(g_1 v\right)^2 \left[ 1 + \frac{x^2}{4} + \cdots \right] ,
\label{eq:Wpmass}
\eeqn
and the corresponding eigenvectors are expressed as 
\beqn
W_0^{a \mu}  &=& v_W^0 A_0^{a \mu} + v_W^1 A_1^{a \mu} \nonumber \\
   &=& \left[ 1 - \frac{x^2}{8} + \cdots \right] A_0^{a \mu} 
         + \left[ \frac{x}{2} + \frac{x^3}{16} + \cdots  \right] A_1^{a \mu} ,
\eeqn
\beqn
W_1^{a \mu}  &=& v_{W'}^0 A_0^{a \mu} + v_{W'}^1 A_1^{a \mu} \nonumber \\
   &=&  \left[ -\frac{x}{2} - \frac{x^3}{16} + \cdots  \right] A_0^{a \mu} 
         + \left[ 1 - \frac{x^2}{8} + \cdots \right]A_1^{a \mu} .
\eeqn
Here, $W_0$ and $W_1$ represent the $W$ and $W'$ gauge bosons, respectively.
From Eqs.~(\ref{eq:Wmass}) and (\ref{eq:Wpmass}), we see that the ratio of $M_W^2$
to $M_{W'}^2$ is of the order of $x^2$:
\begin{equation}
\frac{M_W^2}{M_{W'}^2} = \frac{x^2}{4} \left[ 1 + O(x^2) \right].
\end{equation}

\subsection{NG bosons}
One-loop diagrams which contribute to the 
$b \rightarrow s \gamma$ amplitude involve charged NG-boson propagator and 
its couplings to gauge bosons. For the purpose of obtaining expressions for 
these propagator and couplings, we first introduce the following two-by-two 
matrix, 
\beq
Q
 = \frac{v}{\sqrt{2}}
\left(
 \begin{array}{cc}
  g_0 & -g_1 \\
  0 & g_1 
     \end{array}
 \right), 
\eeq
by using which, the mass squared matrix ${\cal M}_W^2$ and its dual 
$\tilde{{\cal M}}_W^2$ can be written as  \cite{Sfetsos:2001qb} 
\beq
{\cal M}_W^2 = Q^T Q,\ \ \ \ \tilde{{\cal M}}_W^2 = Q\, Q^T.
\eeq
$Q$ is diagonalized by the bi-orthogonal rotation, 
\beq
\tilde{R}_w^T Q R_w = Q^{\rm diag} \equiv {\cal M}_W^{\rm diag}, 
\eeq
which leads to 
\beq
({\cal M}_W^2)^{\rm diag} 
= R_w^T {\cal M}_W^2  R_w 
= (Q^{\rm diag})^T (Q^{\rm diag}) 
= ({\cal M}_W^{\rm diag})^2, 
\eeq
and 
\beq
(\tilde{{\cal M}}_W^2)^{\rm diag} 
= \tilde{R}_w^T \tilde{{\cal M}}_W^2  \tilde{R}_w
= (Q^{\rm diag}) (Q^{\rm diag})^T 
= ({\cal M}_W^{\rm diag})^2. 
\eeq
Thus, we conclude that the dual mass matrix $\tilde{{\cal M}}_W^2$ has the same 
eigenvalues as ${\cal M}_W^2$.

The gauge-NG boson mixing term can be obtained by expanding the Lagrangian 
Eq.~(\ref{eq:L}), 
\beq
{\cal L}_{\rm GB}^{\rm mix}\ 
= \ - {{\bf A}_\mu^{a}}^T\, Q^T\, \partial^\mu {\bf \Pi}^a \ 
= \ - {{\bf W}_\mu^{a}}^T\, {\cal M}_W^{\rm diag}\, \partial^\mu \tilde{{\bf \Pi}}^a,
\label{eq:mix}
\eeq
with 
\beq
{\bf A}^{a \mu} = 
 \left(
 \begin{array}{c}
  A_0^{a \mu} \\
  A_1^{a \mu}
     \end{array}
 \right), \ \ \ 
{\bf W}^{a \mu} = 
 \left(
 \begin{array}{c}
  W_0^{a \mu} \\
  W_1^{a \mu}
     \end{array}
 \right), \ \ \ 
{\bf \Pi}^{a} = 
 \left(
 \begin{array}{c}
  \pi_1^a \\
  \pi_2^a
     \end{array}
 \right), \ \ \ 
\tilde{{\bf \Pi}}^{a} = 
 \left(
 \begin{array}{c}
  \tilde{\pi}_0^a \\
  \tilde{\pi}_1^a
     \end{array}
 \right), \ \ \ 
\eeq
where $\{ {\bf W}^{a \mu} \}$ are mass-eigenbasis fields and 
$\{ \tilde{{\bf \Pi}}^{a} \}$ are ``eaten'' NG fields, which are connected 
to the site gauge bosons $\{ {\bf A}^{a \mu} \}$ and link NG bosons 
$\{ {\bf \Pi}^{a} \}$ via 
\beq
{\bf W}^{a \mu} = R_w^T\, {\bf A}^{a \mu},\ \ \ 
\tilde{{\bf \Pi}}^{a} = \tilde{R}_w^T\, {\bf \Pi}^{a}.
\label{eq:Wtrans}
\eeq
Hence, the eigenvector of the ``eaten" NG bosons are determined by 
the eigenstates of the dual mass squared matrix $\tilde{{\cal M}}_W^2$.
The gauge-NG mixing (\ref{eq:mix}) can be removed by the familiar 
$R_\xi$ gauge-fixing term for the charged gauge sector, 
\beq 
{\cal L}_{\rm gf} = \sum_{n=0}^{1} -\frac{1}{2\xi} (F_n^a)^2,\ \ \ \ \ \ 
F_n^a = \partial_\mu W_n^{a \mu} + \xi M_n \tilde{\pi}_n^a,\ \ \ \ \ \ 
(a=1, 2)
\label{eq:gf}
\eeq
where $M_n$ ($n=0, 1$) represents the mass of $n$-th gauge boson $W_n$. 
From Eq.~(\ref{eq:gf}), one finds that the mass of the NG boson ``eaten" by 
the mass-eigenstate gauge boson $W_n^{a \mu}$ is given by 
$M_{\tilde{\pi}_n^a}^2 = \xi M_n^2$.

\vspace{1cm}
\section{Gauge/NG-boson couplings}
\label{app:GNGcouplings}
Here, we summarize the forms of couplings which involve gauge bosons 
and NG bosons, which are needed for the calculation 
of the $b \rightarrow s \gamma$ amplitude.

\subsection{Triple-Gauge-Boson Vertices}
The triple-gauge-boson couplings which involve a photon and two charged 
gauge bosons are expressed as 
\beq
g_{\gamma W_i W_j} = g_0 v_{\gamma}^0 v_{W_i}^0 v_{W_j}^0
 + g_1 v_{\gamma}^1 v_{W_i}^0 v_{W_j}^1.
\eeq
Here, $W_{i}$ and $W_{j}$ represent either $W_0 (\equiv W)$ or $W_1 (\equiv W')$ 
gauge bosons. From the profile of the photon and the orthonormality of the 
eigenvectors for the charged gauge bosons, couplings become
\beq
g_{\gamma W W} = g_{\gamma W' W'} = e,\ \ \ \ 
g_{\gamma W' W} = 0. 
\eeq

\subsection{$\tilde{\pi}_i^+ \tilde{\pi}_j^- \gamma$ Vertices}
Interactions between two NG modes and a gauge boson are obtained by expanding 
the appropriate terms of Eq.~(\ref{eq:L}):
\beq
{\cal L}_{\pi\pi A} = \frac{1}{2} \epsilon^{abc} \sum_{j=1}^{2} \pi_j^a (\partial_\mu \pi_j^b) 
  \left( g_j A_j^{c \mu} + g_{j-1} A_{j-1}^{c \mu} \right), 
\eeq
with $A_{j=2}^{1\, \mu} = A_{j=2}^{2\, \mu} = 0$ is understood.
$\tilde{\pi}_i^+ \tilde{\pi}_j^- \gamma$ vertices are then derived by expressing the above 
interactions in terms of mass-eigenmodes.
\beqn
{\cal L}_{\tilde{\pi} \tilde{\pi} \gamma} &=& 
i e \sum_{j=1}^{2} \sum_{k, \ell=0}^{1} ( \tilde{R}_w )_{j k} ( \tilde{R}_w )_{j \ell} \ 
\left( \tilde{\pi}_k^- \overleftrightarrow{\partial_\mu} \tilde{\pi}_\ell^+ \right) Z_0^\mu, \nonumber \\
  &=&  
  i e 
  \sum_{\ell =0}^{1}
  \left( \tilde{\pi}_\ell^- \overleftrightarrow{\partial_\mu} \tilde{\pi}_\ell^+ \right) 
  Z_0^\mu,
\eeqn
where $Z_0^\mu$ represents the photon field, and $\overleftrightarrow{\partial_\mu}$ is 
defined as 
$(A \overleftrightarrow{\partial_\mu} B ) \equiv A (\partial_\mu B) - (\partial_\mu A) B$. 
In the last step, we used the orthonormal condition of the rotational matrix $\tilde{R}_w$.

\subsection{$W_i^{\pm} \tilde{\pi}_j^{\mp} \gamma$ Vertices}
Couplings among the photon, charged gauge boson and NG mode can be  
obtained by expanding terms in Eq.~(\ref{eq:L}) which involve one NG mode 
and two gauge bosons:
\beq
{\cal L}_{\pi A A} = \frac{v}{\sqrt{2}}  \sum_{j=1}^2
g_{j-1} g_{j}\, \epsilon^{abc}
A_j^{a \mu} A_{j-1 \mu}^{b} \pi_j^c, 
\eeq
with $A_{j=2}^{1\, \mu} = A_{j=2}^{2\, \mu} = 0$ is understood.
$W_i^\pm \tilde{\pi}_j^\mp \gamma$ vertices are then derived by expressing the above 
interactions in terms of mass-eigenmodes.
\beqn
{\cal L}_{W \tilde{\pi} \gamma} &=& 
i e \frac{v}{\sqrt{2}} \sum_{j=1}^{2} \sum_{k, \ell=0}^{1}
\left[ \left\{  g_{j-1} ( R_w )_{j-1,\, k} - g_j ( R_w )_{j k}  \right\}  ( \tilde{R}_w )_{j \ell} \right]
\left( W_{k \mu}^- \tilde{\pi}_{\ell}^+ -W_{k \mu}^+ \tilde{\pi}_{\ell}^-  \right) Z_0^\mu, \nonumber \\
&=& 
i e \sum_{k, \ell=0}^{1}
( R_w^T Q^T \tilde{R}_w )_{k \ell} 
\left( W_{k \mu}^- \tilde{\pi}_{\ell}^+ -W_{k \mu}^+ \tilde{\pi}_{\ell}^-  \right) Z_0^\mu, \nonumber \\   
&=& 
i e \sum_{k, \ell=0}^{1}
( {\cal M}_W^{\rm diag} )_{k \ell} 
\left( W_{k \mu}^- \tilde{\pi}_{\ell}^+ -W_{k \mu}^+ \tilde{\pi}_{\ell}^-  \right) Z_0^\mu, \nonumber \\   
&=& 
i e \sum_{\ell=0}^{1}
M_{W_\ell} 
\left( W_{\ell \mu}^- \tilde{\pi}_{\ell}^+ -W_{\ell \mu}^+ \tilde{\pi}_{\ell}^-  \right) Z_0^\mu.
\eeqn

\vspace{5mm}
\section{Fermion couplings to charged gauge/NG bosons}
\label{app:Fcouplings}
In this appendix, we summarize the forms of fermion couplings to 
charged gauge/NG bosons. The following light/heavy up/down quark 
mass matrices are used in the expressions:
\begin{equation}
M_u = \left(
\begin{array}{ccc}
  m_u & 0 & 0 \\
  0 & m_c & 0 \\
  0 & 0 &  m_t
\end{array}
\right),\ \ \ 
M_d = \left(
\begin{array}{ccc}
  m_d & 0 & 0 \\
  0 & m_s & 0 \\
  0 & 0 &  m_b
\end{array}
\right),
\nonumber
\end{equation}
\begin{equation}
M_U = \left(
\begin{array}{ccc}
  m_U & 0 & 0 \\
  0 & m_C & 0 \\
  0 & 0 &  m_T
\end{array}
\right),\ \ \ 
M_d = \left(
\begin{array}{ccc}
  m_D & 0 & 0 \\
  0 & m_S & 0 \\
  0 & 0 &  m_B
\end{array}
\right).
\nonumber
\end{equation}

\subsection{Left-handed fermion couplings to charged gauge bosons}
\label{app:Lgauge}
\noindent\underline{Left-handed fermion couplings to the $W$ boson}
\begin{equation}
\frac{g_L^W}{\sqrt{2}} W^{+\, \mu}
\left(
\begin{array}{cccccc}
\bar{u}_L & 
\bar{c}_L & 
\bar{t}_L & 
\bar{U}_L & 
\bar{C}_L & 
\bar{T}_L
\end{array}
\right) \gamma_\mu 
         \left( 
           \begin{array}{ccc|ccc}
           &&&&&\\
              & V_L^{(\ell \ell)} & & & V_L^{(\ell h)} & \\ 
           &&&&&\\ \hline
           &&&&&\\
              & V_L^{(h \ell)} & & & V_L^{(h h)} & \\ 
           &&&&&
           \end{array}
         \right)
         \left( 
           \begin{array}{c}
              d_L\\
              s_L\\
              b_L\\
              D_L\\
              S_L\\
              B_L
           \end{array}
         \right) + h.c.
\end{equation}

\beqn
V_L^{(\ell \ell)}
&=& 
\left(
\begin{array}{ccc}
  1 & 0 & 0 \\
  0 & 1 & 0 \\
  0 & 0 &    1+ \frac{\varepsilon_{tR}^2}{4(1+\varepsilon_{tR}^2)^2} x^2 +  O(x^4)
\end{array}
\right)
  V^{(0)}, \\
V_L^{(\ell h)}
&=& 
\frac{x}{2\sqrt{2}}
\left(
\begin{array}{ccc}
  1 & 0 & 0 \\
  0 & 1 & 0 \\
  0 & 0 &    \frac{1+2\varepsilon_{tR}^2}{1+\varepsilon_{tR}^2}  +  O(x^2)
\end{array}
\right)
V^{(0)}, \\
V_L^{(h \ell)}
&=& 
\frac{x}{2\sqrt{2}}
\left(
\begin{array}{ccc}
  1 & 0 & 0 \\
  0 & 1 & 0 \\
  0 & 0 &    \frac{1-\varepsilon_{tR}^2}{1+\varepsilon_{tR}^2}  +  O(x^2)
\end{array}
\right)
V^{(0)}, \\
V_L^{(h h)}
&=& 
\frac{1}{2}\left(
\begin{array}{ccc}
  1+ x^2 +  O(x^4) & 0 & 0 \\
  0 & 1+ x^2 +  O(x^4) & 0 \\
  0 & 0 &    1+ \frac{\varepsilon_{tR}^4+6\varepsilon_{tR}^2+4}{4(1+\varepsilon_{tR}^2)^2} x^2 +  O(x^4)
\end{array}
\right) 
V^{(0)}.
\eeqn

\noindent\underline{Left-handed fermion couplings to the $W'$ boson}
\begin{equation}
\frac{g_L^W}{\sqrt{2}} {W'}^{+\, \mu}
\left(
\begin{array}{cccccc}
\bar{u}_L & 
\bar{c}_L & 
\bar{t}_L & 
\bar{U}_L & 
\bar{C}_L & 
\bar{T}_L
\end{array}
\right) \gamma_\mu 
         \left( 
           \begin{array}{ccc|ccc}
           &&&&&\\
              & {V'}_L^{(\ell \ell)} & & & {V'}_L^{(\ell h)} & \\ 
           &&&&&\\ \hline
           &&&&&\\
              & {V'}^{(h \ell)}_L & & & {V'}_L^{(h h)} & \\ 
           &&&&&
           \end{array}
         \right)
         \left( 
           \begin{array}{c}
              d_L\\
              s_L\\
              b_L\\
              D_L\\
              S_L\\
              B_L
           \end{array}
         \right) + h.c.
\end{equation}

\beqn
{V'}_L^{(\ell \ell)}
&=& 
\left(
\begin{array}{ccc}
  O(\varepsilon_{qR}^2\,  x) & 0 & 0 \\
  0 & O(\varepsilon_{qR}^2\,  x) & 0 \\
  0 & 0 &    - \frac{\varepsilon_{tR}^2}{2(1+\varepsilon_{tR}^2)} x +  O(x^3)
\end{array}
\right)
V^{(0)}, \\
{V'}_L^{(\ell h)}
&=& 
-\frac{1}{\sqrt{2}}
\left(
\begin{array}{ccc}
  1+\frac{3}{8}x^2 + O(x^4) & 0 & 0 \\
  0 & 1+\frac{3}{8}x^2 + O(x^4) & 0 \\
  0 & 0 &    \frac{1}{1+\varepsilon_{tR}^2}  
  + \frac{3+18\varepsilon_{tR}^2+13\varepsilon_{tR}^4+4\varepsilon_{tR}^6}{8(1+\varepsilon_{tR}^2)^3}x^2+ O(x^4)
\end{array}
\right)
V^{(0)}, \nonumber \\ \\
{V'}_L^{(h \ell)}
&=& 
-\frac{1}{\sqrt{2}}
\left(
\begin{array}{ccc}
  1+\frac{3}{8}x^2 + O(x^4) & 0 & 0 \\
  0 & 1+\frac{3}{8}x^2 + O(x^4) & 0 \\
  0 & 0 &    1   
  + \frac{3+6\varepsilon_{tR}^2+\varepsilon_{tR}^4}{8(1+\varepsilon_{tR}^2)^2}x^2+ O(x^4)
  \end{array}
\right) 
V^{(0)}, \\
{V'}_L^{(h h)}
&=& 
\frac{1}{x}\left(
\begin{array}{ccc}
  1-\frac{1}{4} x^2 +  O(x^4) & 0 & 0 \\
  0 & 1- \frac{1}{4}x^2 +  O(x^4) & 0 \\
  0 & 0 &    1- \frac{1}{4(1+\varepsilon_{tR}^2)^2} x^2 +  O(x^4)
\end{array}
\right) V^{(0)} .
\eeqn

\vspace{1cm}
\subsection{Right-handed fermion couplings to charged gauge bosons}
\label{app:Rgauge}
\noindent\underline{Right-handed fermion couplings to the $W$ boson}
\begin{equation}
\frac{g_L^W}{\sqrt{2}} W^{+\, \mu}
\left(
\begin{array}{cccccc}
\bar{u}_R & 
\bar{c}_R & 
\bar{t}_R & 
\bar{U}_R & 
\bar{C}_R & 
\bar{T}_R
\end{array}
\right) \gamma_\mu 
         \left( 
           \begin{array}{ccc|ccc}
           &&&&&\\
              & V_R^{(\ell \ell)} & & & V_R^{(\ell h)} & \\ 
           &&&&&\\ \hline
           &&&&&\\
              & V_R^{(h \ell)} & & & V_R^{(h h)} & \\ 
           &&&&&
           \end{array}
         \right)
         \left( 
           \begin{array}{c}
              d_R\\
              s_R\\
              b_R\\
              D_R\\
              S_R\\
              B_R
           \end{array}
         \right) + h.c.
\end{equation}

\beqn
V_R^{(\ell \ell)}
&=& 
\left(
\begin{array}{ccc}
  \frac{\varepsilon_{uR}^2}{2(1+\varepsilon_{uR}^2)}\frac{1}{m_u} +O(x^2) & 0 & 0 \\
  0 & \frac{\varepsilon_{cR}^2}{2(1+\varepsilon_{cR}^2)}\frac{1}{m_c} +O(x^2) & 0 \\
  0 & 0 &  \frac{\varepsilon_{tR}^2}{2(1+\varepsilon_{tR}^2)}\frac{1}{m_t} +O(x^2)
\end{array}
\right)
V^{(0)}
M_d
, \nonumber\\ \\
V_R^{(\ell h)}
&=& 
\left(
\begin{array}{ccc}
  O(\varepsilon_{uR}) & 0 & 0 \\
  0 & O(\varepsilon_{cR}) & 0 \\
  0 & 0 &    \frac{\varepsilon_{tR}}{2\sqrt{1+\varepsilon_{tR}^2}}  +  O(x^2)
\end{array}
\right)
V^{(0)}, \\
V_R^{(h \ell)}
&=& 
\frac{1}{2}\left(
\begin{array}{ccc}
  \frac{\varepsilon_{tR}/m_t}{\sqrt{1+\varepsilon_{tR}^2}}  +O(x^2)& 0 & 0 \\
  0 & \frac{\varepsilon_{tR}/m_t}{\sqrt{1+\varepsilon_{tR}^2}}  +O(x^2)& 0 \\
  0 & 0 &  \frac{\varepsilon_{tR}/m_t}{(1+\varepsilon_{tR}^2)} +O(x^2)
\end{array}
\right)
V^{(0)}
M_d,   
\nonumber \\
\\
V_R^{(h h)}
&=& 
\frac{1}{2}\left(
\begin{array}{ccc}
  1+ \frac{x^2}{2} +  O(x^4) & 0 & 0 \\
  0 & 1+ \frac{x^2}{2} +  O(x^4) & 0 \\
  0 & 0 &    \frac{1}{\sqrt{1+\varepsilon_{tR}^2}} +  O(x^2)
\end{array}
\right)  
V^{(0)} .
\eeqn
\vspace{1cm}

\vspace{5mm}
\noindent\underline{Right-handed fermion couplings to the $W'$ boson}
\begin{equation}
\frac{g_L^W}{\sqrt{2}} {W'}^{+\, \mu}
\left(
\begin{array}{cccccc}
\bar{u}_R & 
\bar{c}_R & 
\bar{t}_R & 
\bar{U}_R & 
\bar{C}_R & 
\bar{T}_R
\end{array}
\right) \gamma_\mu 
         \left( 
           \begin{array}{ccc|ccc}
           &&&&&\\
              & {V'}_R^{(\ell \ell)} & & & {V'}_R^{(\ell h)} & \\ 
           &&&&&\\ \hline
           &&&&&\\
              & {V'}^{(h \ell)}_R & & & {V'}_R^{(h h)} & \\ 
           &&&&&
           \end{array}
         \right)
         \left( 
           \begin{array}{c}
              d_R\\
              s_R\\
              b_R\\
              D_R\\
              S_R\\
              B_R
           \end{array}
         \right) + h.c.
\end{equation}

\beqn
{V'}_R^{(\ell \ell)}
&=& 
\frac{1}{x}
\left(
\begin{array}{ccc}
 \frac{\varepsilon_{uR}^2}{1+\varepsilon_{uR}^2}\frac{1}{m_u} +O(x^2) & 0 & 0 \\
  0 & \frac{\varepsilon_{cR}^2}{1+\varepsilon_{cR}^2}\frac{1}{m_c} +O(x^2) & 0 \\
  0 & 0 &  \frac{\varepsilon_{tR}^2}{1+\varepsilon_{tR}^2}\frac{1}{m_t} +O(x^2)
\end{array}
\right)
V^{(0)}
M_d, \nonumber\\ \\
{V'}_R^{(\ell h)}
&=& 
\frac{1}{x}
\left(
\begin{array}{ccc}
  O(\varepsilon_{uR}) & 0 & 0 \\
  0 & O(\varepsilon_{cR}) & 0 \\
  0 & 0 &    \frac{\varepsilon_{tR}}{\sqrt{1+\varepsilon_{tR}^2}}  +  O(x^2)
\end{array}
\right)
V^{(0)}, \\
{V'}_R^{(h \ell)}
&=& 
\frac{1}{x}
\left(
\begin{array}{ccc}
  \frac{\varepsilon_{tR}/m_t}{\sqrt{1+\varepsilon_{tR}^2}}  +O(x^2)& 0 & 0 \\
  0 & \frac{\varepsilon_{tR}/m_t}{\sqrt{1+\varepsilon_{tR}^2}} +O(x^2)& 0 \\
  0 & 0 &  \frac{\varepsilon_{tR}/m_t}{(1+\varepsilon_{tR}^2)} +O(x^2)
\end{array}
\right)
V^{(0)}
M_d,\nonumber\\ \\
{V'}_R^{(h h)}
&=& 
\frac{1}{x}\left(
\begin{array}{ccc}
  1+\frac{1}{4} x^2 +  O(x^4) & 0 & 0 \\
  0 & 1+ \frac{1}{4}x^2 +  O(x^4) & 0 \\
  0 & 0 &    \frac{1}{\sqrt{1+\varepsilon_{tR}^2}}  
  + \frac{(1 + 4\varepsilon_{tR}^2 + \varepsilon_{tR}^4 )}  {4(1+\varepsilon_{tR}^2)^{5/2}} x^2 + O(x^4)
\end{array}
\right) V^{(0)}. \nonumber\\
\eeqn

\vspace{1cm}
\subsection{Fermion couplings to the NG bosons} 
\label{app:NG}

\noindent\underline{Fermion couplings to the light NG mode} 

\beqn
&&
- i\frac{\sqrt{2}}{v}  \tilde{\pi}_0^+
\left(
\begin{array}{cccccc}
\bar{u}_L & 
\bar{c}_L & 
\bar{t}_L & 
\bar{U}_L & 
\bar{C}_L & 
\bar{T}_L
\end{array}
\right)  
         \left( 
           \begin{array}{ccc|ccc}
           &&&&&\\
              & \tilde{V}_+^{(\ell \ell)} & & & \tilde{V}_+^{(\ell h)} & \\ 
           &&&&&\\ \hline
           &&&&&\\
              & \tilde{V}_+^{(h \ell)} & & & \tilde{V}_+^{(h h)} & \\ 
           &&&&&
           \end{array}
         \right)
         \left( 
           \begin{array}{c}
              d_R\\
              s_R\\
              b_R\\
              D_R\\
              S_R\\
              B_R
           \end{array}
         \right)  \nonumber \\
&&
- i  \frac{\sqrt{2}}{v}  \tilde{\pi}_0^-
\left(
\begin{array}{cccccc}
\bar{d}_L & 
\bar{s}_L & 
\bar{b}_L & 
\bar{D}_L & 
\bar{S}_L & 
\bar{B}_L
\end{array}
\right)  
         \left( 
           \begin{array}{ccc|ccc}
           &&&&&\\
              & \tilde{V}_-^{(\ell \ell)} & & & \tilde{V}_-^{(\ell h)} & \\ 
           &&&&&\\ \hline
           &&&&&\\
              & \tilde{V}_-^{(h \ell)} & & & \tilde{V}_-^{(h h)} & \\ 
           &&&&&
           \end{array}
         \right)
         \left( 
           \begin{array}{c}
              u_R\\
              c_R\\
              t_R\\
              U_R\\
              C_R\\
              T_R
           \end{array}
         \right)
    + h.c. \nonumber \\
\eeqn

\beqn
\tilde{V}_+^{(\ell \ell)}
&=& 
\left(
\begin{array}{ccc}
  1  & 0 & 0 \\
  0 & 1 & 0 \\
  0 & 0 &  \frac{2+\varepsilon_{tR}^2}{2(1+\varepsilon_{tR}^2)}   
\end{array}
\right)
V^{(0)}
M_d
[1  + O(x^2)]
, \\
\tilde{V}_+^{(\ell h)}
&=& 
\frac{x}{2\sqrt{2}}  \ 
V^{(0)}
M_D
  [1 + O(x^2)], \\
\tilde{V}_+^{(h \ell)}
&=& 
-\frac{1}{\sqrt{2} x}  \ 
V^{(0)} 
M_d
  [1 + O(x^2)], 
\\
\tilde{V}_+^{(h h)}
&=& 
\frac{1}{2}  \ 
V^{(0)} 
\left(
\begin{array}{ccc}
  \varepsilon_{dR}^2 & 0 & 0 \\
  0 & \varepsilon_{sR}^2  & 0 \\
  0 & 0 &   \varepsilon_{bR}^2
\end{array}
\right)
M_D
[1 + O(x^2)].
\eeqn
\beqn
\tilde{V}_-^{(\ell \ell)}
&=& 
V^{(0) \dagger} 
M_u
  \left(
\begin{array}{ccc}
 1-\frac{x^2}{4} +O(x^4) & 0 & 0 \\
  0 & 1-\frac{x^2}{4} +O(x^4) & 0 \\
  0 & 0 &  1-\frac{1+\varepsilon_{tR}^2+\varepsilon_{tR}^4}{4(1+\varepsilon_{tR}^2)^2}x^2 +O(x^4)
\end{array}
\right)\\
\tilde{V}_-^{(\ell h)}
&=& 
\frac{x}{2\sqrt{2}} \,
V^{(0) \dagger}
M_U
\left(
\begin{array}{ccc}
  1 + O(x^2) & 0 & 0 \\
  0 & 1+O(x^2) & 0 \\
  0 & 0 &    \frac{1-\varepsilon_{tR}^2}{1+\varepsilon_{tR}^2} + O(x^2)
\end{array}
\right), \\
\tilde{V}_-^{(h \ell)}
&=& 
-\frac{1}{\sqrt{2} x}\, 
V^{(0) \dagger} 
M_u
\nonumber \\
&& \ 
  \left(
\begin{array}{ccc}
 1-\frac{3x^2}{8} +O(x^4) & 0 & 0 \\
  0 & 1-\frac{3x^2}{8} +O(x^4) & 0 \\
  0 & 0 &  1-\frac{3+6\varepsilon_{tR}^2+5\varepsilon_{tR}^4}{8(1+\varepsilon_{tR}^2)^2}x^2 +O(x^4)
\end{array}
\right) , \\
\tilde{V}_-^{(h h)}
&=& 
\frac{1}{2}\, 
V^{(0) \dagger} 
M_U
  \left(
\begin{array}{ccc}
 O(\varepsilon_{uR}^2) & 0 & 0 \\
  0 & O(\varepsilon_{cR}^2) & 0 \\
  0 & 0 &  \frac{\varepsilon_{tR}^2}{1+\varepsilon_{tR}^2} +O(x^2)
\end{array}
\right).
\eeqn
\vspace{1cm}

\noindent\underline{Fermion couplings to the heavy NG mode} 

\beqn
&&
- i\frac{\sqrt{2}}{v}  \tilde{\pi}_1^+
\left(
\begin{array}{cccccc}
\bar{u}_L & 
\bar{c}_L & 
\bar{t}_L & 
\bar{U}_L & 
\bar{C}_L & 
\bar{T}_L
\end{array}
\right)  
         \left( 
           \begin{array}{ccc|ccc}
           &&&&&\\
              & \tilde{V'}_+^{(\ell \ell)} & & & \tilde{V'}_+^{(\ell h)} & \\ 
           &&&&&\\ \hline
           &&&&&\\
              & \tilde{V'}_+^{(h \ell)} & & & \tilde{V'}_+^{(h h)} & \\ 
           &&&&&
           \end{array}
         \right)
         \left( 
           \begin{array}{c}
              d_R\\
              s_R\\
              b_R\\
              D_R\\
              S_R\\
              B_R
           \end{array}
         \right)  
         \nonumber \\
&&
- i  \frac{\sqrt{2}}{v}  \tilde{\pi}_1^-
\left(
\begin{array}{cccccc}
\bar{d}_L & 
\bar{s}_L & 
\bar{b}_L & 
\bar{D}_L & 
\bar{S}_L & 
\bar{B}_L
\end{array}
\right)  
         \left( 
           \begin{array}{ccc|ccc}
           &&&&&\\
              & \tilde{V'}_-^{(\ell \ell)} & & & \tilde{V'}_-^{(\ell h)} & \\ 
           &&&&&\\ \hline
           &&&&&\\
              & \tilde{V'}_-^{(h \ell)} & & & \tilde{V'}_-^{(h h)} & \\ 
           &&&&&
           \end{array}
         \right)
         \left( 
           \begin{array}{c}
              u_R\\
              c_R\\
              t_R\\
              U_R\\
              C_R\\
              T_R
           \end{array}
         \right) + h.c.
         \nonumber \\
\eeqn

\beqn
\tilde{V'}_+^{(\ell \ell)}
&=& -\frac{1}{2}
\left(
\begin{array}{ccc}
  \frac{\varepsilon_{uR}^2}{1+\varepsilon_{uR}^2}  & 0 & 0 \\
  0 & \frac{\varepsilon_{cR}^2}{1+\varepsilon_{cR}^2} & 0 \\
  0 & 0 &   \frac{\varepsilon_{tR}^2}{1+\varepsilon_{tR}^2} 
\end{array}
\right)
V^{(0)} 
M_d
[1+O(x^2)],  \\
\tilde{V'}_+^{(\ell h)}
&=& - \frac{x}{2\sqrt{2}}\,
V^{(0)} 
M_D
[1+O(x^2)],  \\
\tilde{V'}_+^{(h \ell)}
&=& - \frac{1}{\sqrt{2} x}\,
V^{(0)} 
M_d
[1+O(x^2)],  \\
\tilde{V'}_+^{(h h)}
&=&  \frac{1}{2}\,
V^{(0)} 
\left(
\begin{array}{ccc}
 \varepsilon_{dR}^2& 0 & 0 \\
  0 & \varepsilon_{sR}^2 & 0 \\
  0 & 0 &   \varepsilon_{bR}^2
\end{array}
\right)
M_D
[1+O(x^2)]. 
\eeqn

\beqn
\tilde{V'}_-^{(\ell \ell)}
&=& -\frac{1}{2}
\left(
\begin{array}{ccc}
  \frac{\varepsilon_{dR}^2}{1+\varepsilon_{dR}^2} +O(x^2) & 0 & 0 \\
  0 & \frac{\varepsilon_{sR}^2}{1+\varepsilon_{sR}^2} +O(x^2) & 0 \\
  0 & 0 &   \frac{\varepsilon_{bR}^2}{1+\varepsilon_{bR}^2} +O(x^2)
\end{array}
\right)
V^{(0) \dagger} 
M_u, \\
\tilde{V'}_-^{(\ell h)}
&=& 
-\frac{x}{2\sqrt{2}}  \ 
V^{(0) \dagger}
M_U
[1 + O(x^2)] , \\
\tilde{V'}_-^{(h \ell)}
&=& 
-\frac{1}{\sqrt{2} x}  \ 
V^{(0) \dagger}
M_u
[1 + O(x^2)] ,
\\
\tilde{V'}_-^{(h h)}
&=& 
\frac{1}{2}  \ 
V^{(0) \dagger}
M_U
\left(
\begin{array}{ccc}
  \frac{\varepsilon_{uR}^2}{1+\varepsilon_{uR}^2} + O(x^2) & 0 & 0 \\
  0 & \frac{\varepsilon_{cR}^2}{1+\varepsilon_{cR}^2}  +O(x^2) & 0 \\
  0 & 0 &  \frac{\varepsilon_{tR}^2}{1+\varepsilon_{tR}^2}   + O(x^2)
\end{array}
\right). 
\eeqn

\end{document}